\documentclass[%
 reprint,
superscriptaddress,
 preprintnumbers,
 amsmath,amssymb,
 aps,
 prl,
]{revtex4-1}

\usepackage[colorlinks=true,linkcolor=blue,citecolor=blue, urlcolor=blue]{hyperref}   

\usepackage{graphicx}
\usepackage{dcolumn}
\usepackage{bm}

\usepackage{xcolor}
\def\Fig#1{Fig.~\ref{#1}}
\def\Eq#1{Eq.~\eqref{#1}}
\def\Eqs#1{Eqs.~\eqref{#1}}
\def\eq#1{\eqref{#1}}

\newcommand{\fakesection}[1]{%
  \par\refstepcounter{section}
  \sectionmark{#1}
  \addcontentsline{toc}{section}{\protect\numberline{\thesection}#1}
  \textbf{#1.}
}

\begin{document}

\preprint{CERN-TH-2020-126}

\title{Discovering Partonic Rescattering in Light Nucleus Collisions}
\author{Alexander Huss}
\email{alexander.huss@cern.ch}
\affiliation{Theoretical Physics Department, CERN, CH-1211 Gen\`eve 23, Switzerland}
\author{Aleksi Kurkela}
 \email{a.k@cern.ch}
\affiliation{Theoretical Physics Department, CERN, CH-1211 Gen\`eve 23, Switzerland}
 \affiliation{Faculty of Science and Technology, University of Stavanger,  4036 Stavanger, Norway}
\author{Aleksas Mazeliauskas}
\email{aleksas.mazeliauskas@cern.ch}
\affiliation{Theoretical Physics Department, CERN, CH-1211 Gen\`eve 23, Switzerland}
\author{Risto Paatelainen}
\email{risto.sakari.paatelainen@cern.ch}
\affiliation{Theoretical Physics Department, CERN, CH-1211 Gen\`eve 23, Switzerland}
\author{Wilke van der Schee}
\email{wilke.van.der.schee@cern.ch}
\affiliation{Theoretical Physics Department, CERN, CH-1211 Gen\`eve 23, Switzerland}
\author{Urs Achim Wiedemann}
 \email{urs.wiedemann@cern.ch}
\affiliation{Theoretical Physics Department, CERN, CH-1211 Gen\`eve 23, Switzerland}

\date{\today}

\begin{abstract}
We demonstrate that oxygen-oxygen collisions at the LHC provide unprecedented sensitivity to parton energy loss 
in a system whose size is comparable to those created in very peripheral heavy-ion collisions.
With leading and next-to-leading order calculations of nuclear modification factors, we show that the baseline in the absence of partonic rescattering is known with up to 2\% theoretical accuracy in inclusive oxygen-oxygen collisions.
Surprisingly, a $Z$-boson normalized nuclear modification factor does not lead to higher theoretical accuracy 
within current uncertainties of nuclear parton distribution functions.
We study a broad range of  parton energy loss models and we find that the expected signal of partonic rescattering can be disentangled from the baseline by measuring charged hadron spectra in the range $20\,\text{GeV}<p_T<100\,\text{GeV}$. 
\end{abstract}

\maketitle

\fakesection{Introduction} Evidence for the formation of deconfined QCD matter---the quark-gluon plasma (QGP)---in nucleus-nucleus (AA) collisions at the LHC and at RHIC comes from several classes of experimental signatures: the suppression of high-momentum hadronic yields (\emph{parton energy loss}), the momentum anisotropy seen in multi-particle correlations (\emph{collective flow}), 
 the increased fraction of strange hadron yields (\emph{strangeness enhancement}), the exponential spectra of electromagnetic probes (\emph{thermal radiation}), and others~\cite{Adcox:2001jp, Adler:2002xw, Aamodt:2010jd, CMS:2012aa, Aad:2015wga, Abelev:2014mda, Aaboud:2017acw, ABELEV:2013zaa, Abelev:2013xaa,Adam:2015lda}. Several of these findings signal the presence of partonic rescattering in the QCD medium produced in AA collisions. 
Even in smaller collision systems, in which interactions may be so feeble that the systems evolve close to free streaming, a smaller but nonvanishing strength of these signatures is expected.

\begin{figure}
  \centering
  \includegraphics[width=\linewidth]{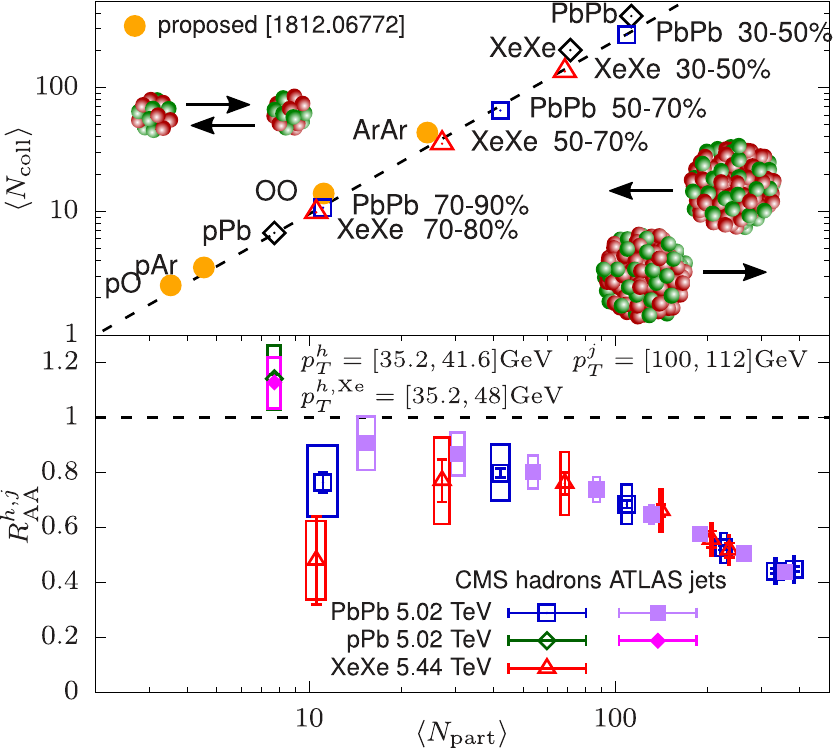}
    \caption{(top) The number of binary collisions as a function of 
    participant nucleons in minimum bias nucleus-nucleus, proton-nucleus and centrality selected heavy-ion collisions. (bottom) Measured hadron and jet nuclear modification factors in PbPb, XeXe and pPb collisions~\cite{Khachatryan:2016odn,Sirunyan:2018eqi, Aaboud:2018twu,ATLAS:2014cpa}. 
    Error bars are statistical, while boxes are the combined systematic, luminosity, and  $\left<T_\text{AA}\right>$ uncertainties.  $\left<T_\text{AA}\right>$ uncertainty dominates in peripheral AA collisions.
    }
    \label{fig1}
\end{figure}

Much experimental effort at the LHC has gone recently into characterizing
emergent QCD medium properties
as a function of the size of the collision system. 
Strangeness enhancement and collective flow have been observed in the most peripheral AA collisions, as well as in proton-nucleus ($p$A) and in proton-proton ($pp$) collisions~\cite{Adam:2015vsf, ALICE:2017jyt, Khachatryan:2015waa, Khachatryan:2016txc}. In marked contrast, no sign of parton energy loss has been observed within current measurement uncertainties in $p$A collisions, and measurements in peripheral AA remain inconclusive because of large systematic uncertainties  (see \Fig{fig1}). However, all parton energy loss models predict some (possibly small) signal in small collision systems. The experimental testing of this robust prediction is arguably one of the most important challenges of the future experimental heavy-ion programs~\cite{Citron:2018lsq, Adolfsson:2020dhm}.

In this Letter, we show how oxygen-oxygen (OO) collisions at the LHC provide
a unique opportunity to discover (small) medium induced energy loss in small systems.

\fakesection{Nuclear modification factor}
The main signal for parton energy loss is the observed suppression of energetic particles in AA collisions. It is typically quantified by the nuclear modification factor 
\begin{align}
  R^{h,j}_\text{AA}(p_T,y)= 
 \frac{1}{\langle T_\text{AA} \rangle} \frac{
   (1/N_\text{ev})~{d}N^{h,j}_\text{AA}/{d}p_T {d}y}{{d}\sigma_{pp}^{h,j}/{d}p_T {d}y}\,,
\label{RAA}
\end{align}
which compares the differential yield in AA 
collisions to the yield in an {\it equivalent number} $\langle N_{\rm coll}\rangle =
\sigma_{pp}^{\rm inel} \, \langle T_\text{AA} \rangle$ of $pp$ collisions. Here, 
$\sigma_{pp}^{\rm inel}$ is the total inelastic $pp$ cross section,  
$\langle T_\text{AA} \rangle$ is the nuclear overlap function within a given centrality interval, and $N_\text{ev}$ is the number of collision events in this centrality interval. ${d}N^{h,j}_\text{AA}/{d}p_T {d}y$ is the differential yield of charged hadrons ($h$) or calorimetrically defined jets ($j$) 
produced in AA collisions at transverse momentum $p_T$ and longitudinal rapidity $y$, and ${d}\sigma_{pp}^{h,j}/{d}p_T {d}y$ is the corresponding differential $pp$ cross section. 

The system size dependence of parton energy loss is typically studied in terms of the centrality dependence of $R_\text{AA}(p_T, y)$. Experimentally,  centrality is defined as the selected percentage of the highest multiplicity events of the total inelastic AA cross section. Theoretically, it is related by Glauber-type models  to $\langle T_\text{AA} \rangle$, to the mean number of participating nucleons $\left<N_\text{part}\right>$ and to the
mean number of nucleon-nucleon collisions $\left<N_\text{coll}\right>$~\cite{Glauber:1970jm, dEnterria:2003xac,Miller:2007ri,Morsch:2017brb}. 
As seen from the top panel of \Fig{fig1}, inclusive (i.e., centrality averaged) OO collisions probe the system size corresponding to 
highly peripheral lead-lead (PbPb) and xenon-xenon (XeXe) collisions.

The differential cross section $d\sigma^{h,j}_{pp}$ entering \Eq{RAA} can be measured precisely and 
it can be calculated at sufficiently high $p_T$ with controlled accuracy in QCD perturbation theory. 
However, the nuclear overlap function $\langle T_\text{AA} \rangle$ depends on the soft physics of total inelastic $pp$ cross section and on the model dependent estimation of binary nucleon-nucleon collisions. 
Estimates of the uncertainties associated to $\langle T_\text{AA} \rangle$ range from $3\%$ in central to $15 \%$ in the
peripheral PbPb collisions~\cite{Khachatryan:2016odn}. In  addition, there are known event selection and geometry biases that in peripheral AA collisions complicate the model 
comparison of nuclear modification factors~\cite{Morsch:2017brb}.
In this way, the characterization of a high-momentum transfer process
becomes dependent on the modeling of  low-energy physics whose uncertainties are difficult to estimate and to improve.
This limits the use of \Eq{RAA} for 
characterizing numerically small medium modifications in very peripheral heavy-ion and $p$A collisions.
A centrality averaged measurement of \Eq{RAA} in OO collisions would have 
a smaller $\langle T_\text{AA} \rangle$ uncertainty than $15\%$, but soft physics assumptions remain~\cite{Eskola:2020lee}.

It is of interest to characterize parton energy loss in the range of $\left<N_\text{part}\right> \sim 10$ 
with measurements independent of soft physics assumptions. 
The study of inclusive, minimum bias $R^{h,j}_\text{AA}$ in collisions of light nuclei allows for this since 
\begin{align}
 R^{h,j}_\text{AA, min bias}(p_T,y)= 
 \frac{1}{A^2} \frac{ {d\sigma^{h,j}_\text{AA}}/{dp_T dy}}{ {d\sigma_{pp}^{h,j}}/{dp_T dy}} 
\label{RAAminbias}
\end{align}
is independent of $\langle T_\text{AA} \rangle$. 
The system size is controlled by selecting nuclei with different nucleon number $A$. Proposed light-ion
collisions with oxygen $A=16$ and argon $A=40$ at the LHC provide a system size scan in the physically interesting
region, see \Fig{fig1}.

\fakesection{Perturbative benchmark calculations}
The ability to discover a small signal of high-$p_T$ partonic rescattering via \Eq{RAAminbias} is now free from soft physics assumptions. It  depends solely on the experimental precision of 
the measurement and on the accuracy with which theory can calculate the null hypothesis, i.e., the value of 
$ R^{h,j}_\text{AA, min bias}$ in the absence of partonic rescattering. 
This null hypothesis depends only on high-momentum transfer processes that can be computed with systematically improvable
accuracy in collinearly factorized perturbative QCD. 
To determine the null hypothesis, we calculate inclusive jet cross section in $pp$ and OO collisions at $\sqrt{s_{NN}} = 7\,\text{TeV}$
as the convolution 
of incoming parton distribution functions (PDFs) with hard matrix elements with the \texttt{NNLOJET} framework~\cite{Currie:2016bfm,Gehrmann:2018szu} and using \texttt{APPLfast} interpolation tables~\cite{Britzger:2019kkb}.
For $pp$ collisions, cross section calculations provide quantitatively reliable predictions at next-to-leading order (NLO) and have been pushed 
to NNLO accuracy or even beyond for many important processes.
For nuclei, the nuclear modifications of the PDFs (nPDFs)
are currently available up to NLO accuracy,
 so we restrict calculations of \Eq{RAAminbias} up to this order. 

\begin{figure}
  \centering
\includegraphics[width=\linewidth]{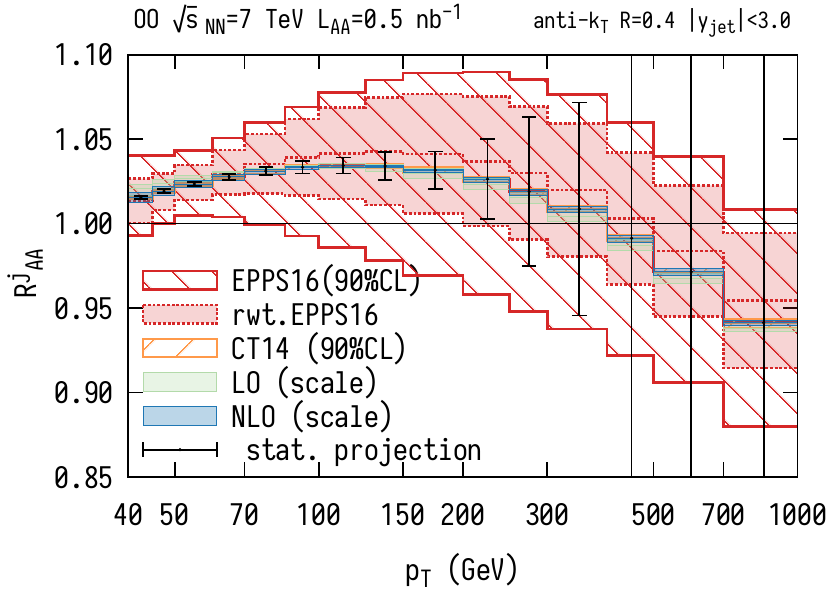}%
  \caption{Minimum bias jet nuclear modification factor \Eq{RAAminbias} for OO collisions in the absence of parton rescattering.  
  The red bands show nPDF 90\% confidence level (CL) (reweighting is done
  by including additional $p$Pb dijet data). 
  Proton PDF (orange) and scale (green and blue) uncertainties are fully correlated and cancel. 
Error bars illustrate statistical  uncertainties for OO mock data at 100\% efficiency (see text for other uncertainties).
}
  \label{fig2}
\end{figure}

Results for the minimum bias nuclear modification factor of jets are shown in \Fig{fig2}. The uncertainties in the proton PDFs 
and in the fixed-order perturbative calculation were estimated using the free proton PDF sets provided by CT14~\cite{Dulat:2015mca}
and by independently varying the factorization and renormalization scales by factors $\tfrac{1}{2}$ and 2 while imposing $\tfrac{1}{2}\leq \mu_\text{R}/\mu_\text{F} \leq 2$. For leading order (LO) and NLO calculations, these theoretical uncertainties 
enter the numerator and denominator of \Eq{RAAminbias} and are found to cancel to a large extent in the ratio.
We checked  that parton-shower (PS) and hadronization effects also largely cancel using the NLO+PS implementation of \texttt{POWHEG}+\texttt{Pythia8}~\cite{Alioli:2010xa}.

Uncertainties of nuclear modification of the free proton PDFs,
however, enter only in the numerator
of \Eq{RAAminbias}. They were calculated using nPDF sets from EPPS16 global fit
including a subset of LHC data on electroweak boson and dijet production in $p$Pb~\cite{Eskola:2016oht}. 
nPDFs constitute the largest
theoretical uncertainty, increasing from $\sim 2\%$ at $p_T = 50\,\text{GeV}$ to   $\sim 7\%$ for $p_T > 200\,\text{GeV}$. Compared to a
conservative 15\% uncertainty estimate on the modeling of  $\langle T_\text{AA} \rangle$ for very peripheral heavy-ion collisions, they are approximately 4 times smaller for $p_T < 100\,\text{GeV}$.
 Moreover,
nPDF uncertainties can be reduced by including additional LHC data. We show this by reweighting nPDF uncertainties with CMS dijet data~\cite{Sirunyan:2018qel} (following the work
of Ref.~\cite{Paukkunen:2014zia, Eskola:2019dui}, see the Supplemental Material).
The nPDF 90\% confidence level band in \Fig{fig2} then shrinks to 1\% (4\%) at low (high) $p_T$, respectively.
This demonstrates that the null hypothesis in the absence of parton energy loss is known with much higher accuracy from \Eq{RAAminbias} than from the centrality dependent measurements of \Eq{RAA}.

To gain insight into whether this higher theoretical accuracy can be exploited in an upcoming OO run, 
we have overlaid in \Fig{fig2} statistical uncertainties of OO mock data  for an integrated luminosity of $\mathcal{L}_\text{AA}=0.5\,\text{nb}^{-1}$ corresponding to a few hours of stable beam  in the ``moderately optimistic'' running scenario of Ref.~\cite{Citron:2018lsq}. The errors displayed on the mock
data do not account for several sources of experimental uncertainties that can only be determined with detailed knowledge of 
the detectors and the machine. 
There are indications that the systematic experimental uncertainties entering \Eq{RAAminbias} can be brought 
down to less than 4\% in the measurement of the jet nuclear modification factor \cite{Aaboud:2018twu}. 
In addition, a precise determination of \Eq{RAAminbias} requires controlling the OO and $pp$ beam luminosities with comparable accuracy~\cite{ATLAS-CONF-2019-021,ATLAS-CONF-2020-010}.
In this case, both the experimental precision and theoretical accuracy of the 
no-parton-energy-loss baseline of \Eq{RAAminbias} in OO would  be high enough to provide unprecedented sensitivity for the search of parton energy loss in systems with  $\left<N_\text{part}\right> \sim 10$.

\begin{figure}
  \centering
  \includegraphics[width=\linewidth]{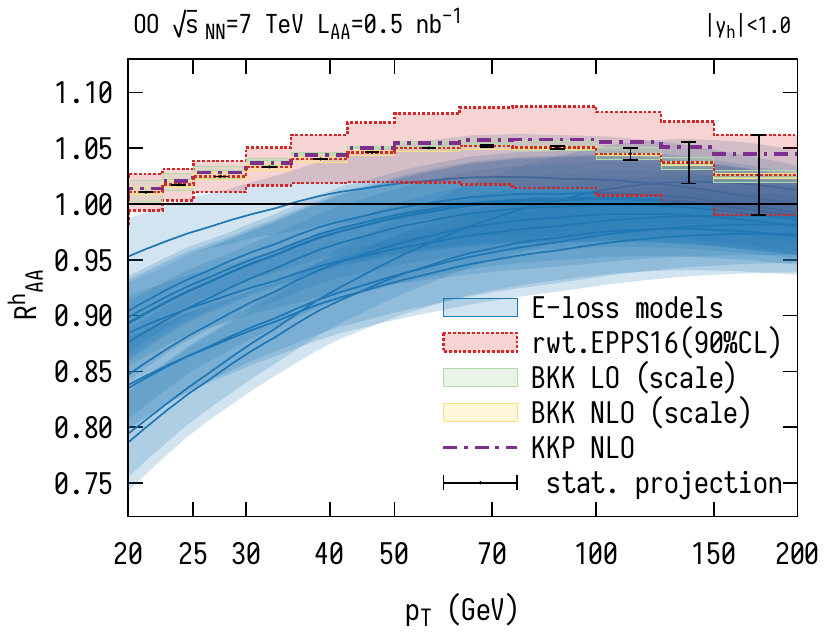}
  \caption{
Minimum bias hadron nuclear modification factor, \Eq{RAAminbias}, for OO collisions.  
A broad range of parton energy loss model predictions (blue bands)~\cite{Huss:2020whe} is overlaid with the baseline  in the absence of parton rescattering.
  The red band shows a reweighted nPDF 90\% confidence level (reweighting is done
  by including additional $p$Pb dijet data). 
  Proton PDF (not shown), scale (green and yellow) uncertainties are fully correlated and cancel. Dot-dashed line shows central NLO prediction with KKP FFs.
Error bars illustrate statistical  uncertainties for OO mock data at 100\% efficiency.
  }
  \label{fig3}
\end{figure}

In close analogy, we have also calculated the nuclear modification factor \Eq{RAAminbias} for single inclusive charged hadron spectra at LO and NLO.
We convoluted the parton spectra with Binnewies-Kniehl-Kramer (BKK)~\cite{Binnewies:1994ju} and Kniehl-Kramer-Potter (KKP)~\cite{Kniehl:2000fe} fragmentation functions (FFs) using the
\texttt{INCNLO} program~\cite{Aversa:1988vb}\footnote{\url{http://lapth.cnrs.fr/PHOX_FAMILY/readme_inc.html}} modified to use LHAPDF grids~\cite{Buckley:2014ana}. We obtained hadronic FFs by summing pion and kaon FFs for BKK and pion, kaon and proton FFs for KKP. We checked that BKK FFs (our default choice) provide a reasonable description of the measured charged hadron spectra at $\sqrt{s}=7\,\text{TeV}$ $pp$ collisions.
In the absence of final state rescattering
in the QCD medium the same FFs enter the numerator and the denominator in \Eq{RAAminbias},
such that the ratio is largely insensitive to the specific choice of FFs, as shown in \Fig{fig3}.
The remaining uncertainty is dominated again by our current knowledge
of nPDFs. As parton fragmentation softens hadron distributions, 
the region of  small $\sim 2\%$ uncertainty lies at a $p_T$ that is shifted compared to the $p_T$ dependence in \Fig{fig2}.

\fakesection{Predictions of parton energy loss}
  The sizable azimuthal momentum anisotropies $v_n$ observed in systems of $\left<N_\text{part}\right> \sim 10$ are interpreted in terms of interactions in the QCD medium. Therefore, qualitatively, some parton energy loss in OO collisions is expected. However, quantitative theoretical expectations for $R_{\text{AA},\text{min bias}}^h$  are model dependent, and there is no \emph{a priori} reason that the effect is large.
The medium modifications of the multiparticle final states giving rise to jets are more
complicated to model than single inclusive hadron spectra, and none of the Monte Carlo tools developed to this end (see, e.g.,~\cite{Zapp:2013vla, Putschke:2019yrg, Schenke:2009gb}) have been tuned to very small collision systems. For these reasons,
we restrict the following discussion of quantitative model expectations for parton energy loss in OO to single inclusive hadron
spectra.

In general, models of parton energy loss supplement the framework of collinearly factorized QCD with assumptions about the
rescattering and ensuing modifications of the final state parton shower in the QCD medium. 
For leading hadron spectra, 
the hard matrix elements are typically convoluted with quenching weights that characterize the parton energy 
loss of the leading parton in the QCD medium prior to hadronization in the vacuum. First perturbative calculations of this parton
rescattering within QCD go back to the works of Baier-Dokshitzer-Mueller-Peigne-Schiff and Zakharov~\cite{Baier:1996kr, Baier:1996sk, Zakharov:1996fv, Zakharov:1997uu} and many others~\cite{Wiedemann:2000za, Gyulassy:2000er, Wang:2001ifa}. Within this framework, a large number of models were developed for the description of $R_\text{AA}^h$ over the last two decades~\cite{Armesto:2011ht}. 
These models
differ in their assumptions about the strength of the rescattering (typically parameterized in terms of the quenching parameter 
$\hat q$ or an equivalent parameter), the time evolution of the medium, the path length dependence, and other details. 
To the best of our knowledge,
none of these models have been used to make predictions for $R_{\text{AA, min bias}}^h$ in OO collisions.

In a companion paper~\cite{Huss:2020whe}, we therefore derive predictions for $R_\text{AA,min bias}^h$ in OO collisions.
This is done by 
building a 
simple modular version of the factorized perturbative QCD framework supplemented with 
parton energy loss.
We have systematically tested the resulting $R_{\text{AA, min bias} }^h(p_T)$ for a wide set of model 
assumptions. All models
were tuned to experimental data
of $R_{\text{AA, min bias} }^h(p_T)$ in  $\sqrt{s_{NN}}=5.02\,\text{TeV}$ PbPb collisions at $p_T \sim 50\,\text{GeV}$~\cite{Khachatryan:2016odn}.
We then predict the $p_T$ and system size dependence.
Although our procedure is not the same as reproducing the various published parton energy loss models (the different model assumptions are embedded all in the same simple setup), we expect that this
characterizes reasonably well the spread in model predictions
for OO collisions. Referring for details to the companion paper~\cite{Huss:2020whe}, we show the final result in \Fig{fig3}.
The blue lines result from overlaying predictions for different modeling assumptions  and thus presents a robust expectation for 
parton energy loss. The blue bands represent model and (reweighted) nPDF uncertainties added in quadrature.
We conclude that a 15\% 
uncertainty in modeling of $\left< T_\text{AA}\right>$ in very 
peripheral PbPb collisions would prevent separating a large fraction of the model predictions  from the null hypothesis.
However, the much improved theoretical accuracy of 
\Eq{RAAminbias} (error bands in \Fig{fig3}) allows for this 
separation for the large majority of models in the range of $20\,\text{GeV}< p_T < 50\,\text{GeV}$,
and for some in the range up to 100 GeV.

\begin{figure}
    \centering
\includegraphics[width=\linewidth]{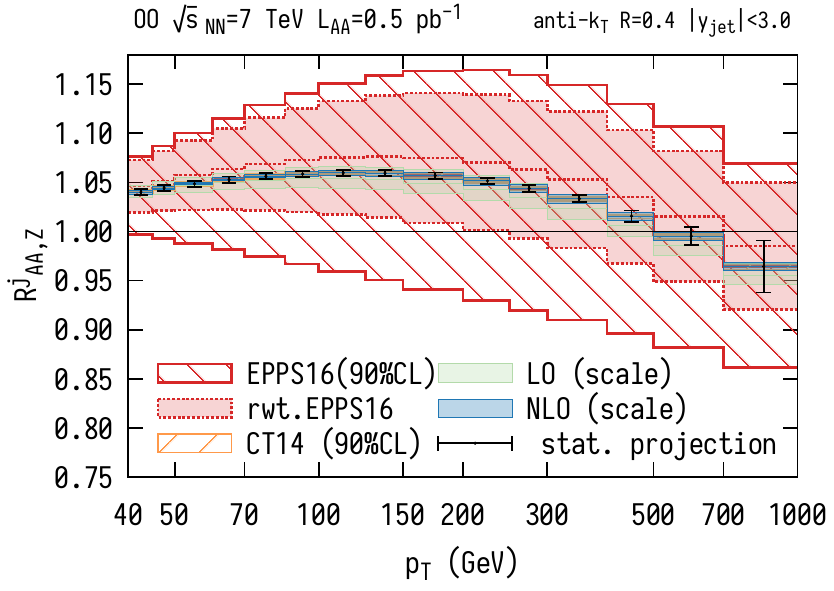}
    \caption{(a) The $Z$-boson normalized jet nuclear modification factor, \Eq{ZRAA},
    for OO collisions in the absence of parton rescattering (analogous to \Fig{fig2}).
    The surprising increase in the red band is due to 
    the anticorrelation of $Z$ and jet nPDF uncertainties (see text).
    The error bars represent statistical uncertainties of OO mock data at 100\% efficiency at an integrated luminosity of ${\mathcal L}_\text{AA}=0.5\,\text{pb}^{-1}$.
    }
    \label{fig4}
\end{figure}

{\bf Opportunities of $Z$-boson measurements.}
While our model studies indicate that the theoretical accuracy  will be sufficient
to discover partonic rescattering in small systems, the use of \Eq{RAAminbias} could potentially be
limited by beam luminosity uncertainties. $Z$-boson production has been long touted as a golden channel 
to measure precisely the hard partonic luminosity~\cite{Dittmar:1997md,Khoze:2000db}.
Therefore,
we consider the $Z$-boson normalized nuclear modification factor
\begin{align}
 R^{h,j}_{\text{AA},Z}(p_T,y)= 
    \frac{\sigma^Z_{pp}}{\sigma^{Z}_\text{AA}} \frac{ {d\sigma^{h,j}_\text{AA}}/{dp_T dy}}{ {d\sigma_{pp}^{h,j}}/{dp_T dy}}\, .
\label{ZRAA}
\end{align}
In comparison to \Eq{RAAminbias}, this measurement has the additional advantage
of the beam luminosity uncertainties canceling in the double ratio of 
cross sections.

OO collisions at LHC can reach an order of magnitude larger effective nucleon-nucleon luminosity than PbPb collisions~\cite{Citron:2018lsq}. A sample of $\mathcal{O}(10^5)$ $Z$ bosons can be
recorded with an integrated  luminosity ${\mathcal L}_{AA}=0.5\,\text{pb}^{-1}$ of OO collisions which corresponds 
nominally to $\mathcal{O}(1 \text{ day})$ stable  running at LHC. This would bring the statistical uncertainties of the normalization in \Eq{ZRAA} below $1\%$.

As both jet and $Z$-boson yields are proportional to the incoming parton flux, we expected that the nPDF uncertainties would also largely cancel in the double ratio. 
In \Fig{fig4} we show the baseline calculation of \Eq{ZRAA} obtained in the same \texttt{NNLOJET} framework and displayed with the same breakdown of theoretical uncertainties as \Fig{fig2}.
The comparison of Figs.~\ref{fig2} and \ref{fig4} makes it clear that our initial assumption was wrong
and that the nPDF uncertainties in \Eq{ZRAA} are larger than those in \Eq{RAAminbias}. The reason for this
is that the $Z$-boson and jet cross sections probe different Bjorken-$x$ ranges and that the nPDF uncertainties of these
ranges turn out to be \emph{anticorrelated} (see the Supplemental Material). 
We conclude that the theoretical accuracy of $Z$-boson normalized nuclear modification factor, \Eq{ZRAA}, relies on a precise knowledge of nPDFs. 
As more LHC data on AA and $p$A collision will be included in the nPDF fits, nPDF uncertainties will be reduced. It would be interesting to study to what extent future  $p$O and OO runs at LHC can improve the current nPDF uncertainties.

\fakesection{Summary} We have started from the observation that the current characterization of parton energy loss in small systems relies
on \emph{centrality dependent} measurements whose construction depends on assumptions about soft physics (in particular manifest in $\langle T_\text{AA} \rangle$).
The associated uncertainties are difficult to improve systematically and they constitute a significant limitation for high precision
measurements of small parton energy loss effects in small collision systems. We have demonstrated with LO and NLO
calculations of the baseline of negligible parton energy loss that
theoretical uncertainties for \emph{inclusive} measurements of nuclear modification factors  are much smaller and as low as 2\% in the kinematically most favorable regions. 
Moreover, these uncertainties can be systematically improved with new data that constrain nPDFs.

We reemphasize that partonic 
rescattering is a prerequisite for quark-gluon plasma formation and  that partonic rescattering is a direct logical consequence of the standard interpretations of azimuthal anisotropies $v_n$ in terms of final state interactions. The possibility that $v_n$ is observed while partonic scattering is absent contradicts 
such phenomenological interpretation of heavy-ion data. The discovery of parton energy loss in small collision systems is therefore
one of the most important challenges of the future experimental heavy-ion program. 
Here, we have shown that the improved
theoretical uncertainty in the baseline calculation of inclusive hadron spectra is needed to separate 
unambiguously model predictions of partonic rescattering from the null hypothesis in the small OO collision system. 
The integrated luminosity to make this possible is $\mathcal {O}(1\, \text{nb}^{-1})$.
Measurements of $Z$-boson normalized $R_{\text{AA},Z}$ would
provide an alternative characterization of parton energy loss in OO collisions that has comparable accuracy and that has the
advantage of the luminosity uncertainties canceling. 
Such measurement require an integrated luminosity of $\mathcal{O}(1\, \text{pb}^{-1})$.
We hope that our proposal helps to 
clarify one of the main outstanding questions in the LHC heavy-ion program and that it 
informs the ongoing discussions about the integrated luminosity required to exploit the unique opportunities of an OO run 
at the LHC.

{\bf Acknowledgements.}
We thank
Rabah Abdul Khalek,
Julien Baglio,
Valerio Bertone, 
Roderik Bruce,
Silvia Ferrario Ravasio,
Oscar Garcia-Montero,
Marc Andre Jebramcik,
John Jowett,
Yen\nobreakdash-Jie Lee,
Andreas Morsch, 
Dennis Perepelitsa, 
Juan Rojo,
Marta Verweij, and
Eleni Vryonidou
for valuable discussions.
\\ 
We thank Hannu Paukkunen for providing EPPS16 parton distribution functions for oxygen, Valerio Bertone for sharing the modified \texttt{INCNLO} code and 
Guilherme Milhano for numerous discussions throughout the project.
\bibliography{master.bib}

\begin{thebibliography}{55}%
\makeatletter
\providecommand \@ifxundefined [1]{%
 \@ifx{#1\undefined}
}%
\providecommand \@ifnum [1]{%
 \ifnum #1\expandafter \@firstoftwo
 \else \expandafter \@secondoftwo
 \fi
}%
\providecommand \@ifx [1]{%
 \ifx #1\expandafter \@firstoftwo
 \else \expandafter \@secondoftwo
 \fi
}%
\providecommand \natexlab [1]{#1}%
\providecommand \enquote  [1]{``#1''}%
\providecommand \bibnamefont  [1]{#1}%
\providecommand \bibfnamefont [1]{#1}%
\providecommand \citenamefont [1]{#1}%
\providecommand \href@noop [0]{\@secondoftwo}%
\providecommand \href [0]{\begingroup \@sanitize@url \@href}%
\providecommand \@href[1]{\@@startlink{#1}\@@href}%
\providecommand \@@href[1]{\endgroup#1\@@endlink}%
\providecommand \@sanitize@url [0]{\catcode `\\12\catcode `\$12\catcode
  `\&12\catcode `\#12\catcode `\^12\catcode `\_12\catcode `\%12\relax}%
\providecommand \@@startlink[1]{}%
\providecommand \@@endlink[0]{}%
\providecommand \url  [0]{\begingroup\@sanitize@url \@url }%
\providecommand \@url [1]{\endgroup\@href {#1}{\urlprefix }}%
\providecommand \urlprefix  [0]{URL }%
\providecommand \Eprint [0]{\href }%
\providecommand \doibase [0]{http://dx.doi.org/}%
\providecommand \selectlanguage [0]{\@gobble}%
\providecommand \bibinfo  [0]{\@secondoftwo}%
\providecommand \bibfield  [0]{\@secondoftwo}%
\providecommand \translation [1]{[#1]}%
\providecommand \BibitemOpen [0]{}%
\providecommand \bibitemStop [0]{}%
\providecommand \bibitemNoStop [0]{.\EOS\space}%
\providecommand \EOS [0]{\spacefactor3000\relax}%
\providecommand \BibitemShut  [1]{\csname bibitem#1\endcsname}%
\let\auto@bib@innerbib\@empty
\bibitem [{\citenamefont {Adcox}\ \emph {et~al.}(2002)\citenamefont {Adcox}
  \emph {et~al.}}]{Adcox:2001jp}%
  \BibitemOpen
  \bibfield  {author} {\bibinfo {author} {\bibfnamefont {K.}~\bibnamefont
  {Adcox}} \emph {et~al.} (\bibinfo {collaboration} {PHENIX}),\ }\href
  {\doibase 10.1103/PhysRevLett.88.022301} {\bibfield  {journal} {\bibinfo
  {journal} {Phys. Rev. Lett.}\ }\textbf {\bibinfo {volume} {88}},\ \bibinfo
  {pages} {022301} (\bibinfo {year} {2002})},\ \Eprint
  {http://arxiv.org/abs/nucl-ex/0109003} {arXiv:nucl-ex/0109003} \BibitemShut
  {NoStop}%
\bibitem [{\citenamefont {Adler}\ \emph {et~al.}(2002)\citenamefont {Adler}
  \emph {et~al.}}]{Adler:2002xw}%
  \BibitemOpen
  \bibfield  {author} {\bibinfo {author} {\bibfnamefont {C.}~\bibnamefont
  {Adler}} \emph {et~al.} (\bibinfo {collaboration} {STAR}),\ }\href {\doibase
  10.1103/PhysRevLett.89.202301} {\bibfield  {journal} {\bibinfo  {journal}
  {Phys. Rev. Lett.}\ }\textbf {\bibinfo {volume} {89}},\ \bibinfo {pages}
  {202301} (\bibinfo {year} {2002})},\ \Eprint
  {http://arxiv.org/abs/nucl-ex/0206011} {arXiv:nucl-ex/0206011} \BibitemShut
  {NoStop}%
\bibitem [{\citenamefont {Aamodt}\ \emph {et~al.}(2011)\citenamefont {Aamodt}
  \emph {et~al.}}]{Aamodt:2010jd}%
  \BibitemOpen
  \bibfield  {author} {\bibinfo {author} {\bibfnamefont {K.}~\bibnamefont
  {Aamodt}} \emph {et~al.} (\bibinfo {collaboration} {ALICE}),\ }\href
  {\doibase 10.1016/j.physletb.2010.12.020} {\bibfield  {journal} {\bibinfo
  {journal} {Phys. Lett. B}\ }\textbf {\bibinfo {volume} {696}},\ \bibinfo
  {pages} {30} (\bibinfo {year} {2011})},\ \Eprint
  {http://arxiv.org/abs/1012.1004} {arXiv:1012.1004 [nucl-ex]} \BibitemShut
  {NoStop}%
\bibitem [{\citenamefont {Chatrchyan}\ \emph {et~al.}(2012)\citenamefont
  {Chatrchyan} \emph {et~al.}}]{CMS:2012aa}%
  \BibitemOpen
  \bibfield  {author} {\bibinfo {author} {\bibfnamefont {S.}~\bibnamefont
  {Chatrchyan}} \emph {et~al.} (\bibinfo {collaboration} {CMS}),\ }\href
  {\doibase 10.1140/epjc/s10052-012-1945-x} {\bibfield  {journal} {\bibinfo
  {journal} {Eur. Phys. J. C}\ }\textbf {\bibinfo {volume} {72}},\ \bibinfo
  {pages} {1945} (\bibinfo {year} {2012})},\ \Eprint
  {http://arxiv.org/abs/1202.2554} {arXiv:1202.2554 [nucl-ex]} \BibitemShut
  {NoStop}%
\bibitem [{\citenamefont {Aad}\ \emph {et~al.}(2015{\natexlab{a}})\citenamefont
  {Aad} \emph {et~al.}}]{Aad:2015wga}%
  \BibitemOpen
  \bibfield  {author} {\bibinfo {author} {\bibfnamefont {G.}~\bibnamefont
  {Aad}} \emph {et~al.} (\bibinfo {collaboration} {ATLAS}),\ }\href {\doibase
  10.1007/JHEP09(2015)050} {\bibfield  {journal} {\bibinfo  {journal} {JHEP}\
  }\textbf {\bibinfo {volume} {09}},\ \bibinfo {pages} {050} (\bibinfo {year}
  {2015}{\natexlab{a}})},\ \Eprint {http://arxiv.org/abs/1504.04337}
  {arXiv:1504.04337 [hep-ex]} \BibitemShut {NoStop}%
\bibitem [{\citenamefont {Abelev}\ \emph
  {et~al.}(2014{\natexlab{a}})\citenamefont {Abelev} \emph
  {et~al.}}]{Abelev:2014mda}%
  \BibitemOpen
  \bibfield  {author} {\bibinfo {author} {\bibfnamefont {B.~B.}\ \bibnamefont
  {Abelev}} \emph {et~al.} (\bibinfo {collaboration} {ALICE}),\ }\href
  {\doibase 10.1103/PhysRevC.90.054901} {\bibfield  {journal} {\bibinfo
  {journal} {Phys. Rev. C}\ }\textbf {\bibinfo {volume} {90}},\ \bibinfo
  {pages} {054901} (\bibinfo {year} {2014}{\natexlab{a}})},\ \Eprint
  {http://arxiv.org/abs/1406.2474} {arXiv:1406.2474 [nucl-ex]} \BibitemShut
  {NoStop}%
\bibitem [{\citenamefont {Aaboud}\ \emph {et~al.}(2017)\citenamefont {Aaboud}
  \emph {et~al.}}]{Aaboud:2017acw}%
  \BibitemOpen
  \bibfield  {author} {\bibinfo {author} {\bibfnamefont {M.}~\bibnamefont
  {Aaboud}} \emph {et~al.} (\bibinfo {collaboration} {ATLAS}),\ }\href
  {\doibase 10.1140/epjc/s10052-017-4988-1} {\bibfield  {journal} {\bibinfo
  {journal} {Eur. Phys. J. C}\ }\textbf {\bibinfo {volume} {77}},\ \bibinfo
  {pages} {428} (\bibinfo {year} {2017})},\ \Eprint
  {http://arxiv.org/abs/1705.04176} {arXiv:1705.04176 [hep-ex]} \BibitemShut
  {NoStop}%
\bibitem [{\citenamefont {Abelev}\ \emph
  {et~al.}(2014{\natexlab{b}})\citenamefont {Abelev} \emph
  {et~al.}}]{ABELEV:2013zaa}%
  \BibitemOpen
  \bibfield  {author} {\bibinfo {author} {\bibfnamefont {B.~B.}\ \bibnamefont
  {Abelev}} \emph {et~al.} (\bibinfo {collaboration} {ALICE}),\ }\href
  {\doibase 10.1016/j.physletb.2014.05.052} {\bibfield  {journal} {\bibinfo
  {journal} {Phys. Lett. B}\ }\textbf {\bibinfo {volume} {728}},\ \bibinfo
  {pages} {216} (\bibinfo {year} {2014}{\natexlab{b}})},\ \bibinfo {note}
  {[Erratum: Phys.Lett.B 734, 409--410 (2014)]},\ \Eprint
  {http://arxiv.org/abs/1307.5543} {arXiv:1307.5543 [nucl-ex]} \BibitemShut
  {NoStop}%
\bibitem [{\citenamefont {Abelev}\ \emph {et~al.}(2013)\citenamefont {Abelev}
  \emph {et~al.}}]{Abelev:2013xaa}%
  \BibitemOpen
  \bibfield  {author} {\bibinfo {author} {\bibfnamefont {B.~B.}\ \bibnamefont
  {Abelev}} \emph {et~al.} (\bibinfo {collaboration} {ALICE}),\ }\href
  {\doibase 10.1103/PhysRevLett.111.222301} {\bibfield  {journal} {\bibinfo
  {journal} {Phys. Rev. Lett.}\ }\textbf {\bibinfo {volume} {111}},\ \bibinfo
  {pages} {222301} (\bibinfo {year} {2013})},\ \Eprint
  {http://arxiv.org/abs/1307.5530} {arXiv:1307.5530 [nucl-ex]} \BibitemShut
  {NoStop}%
\bibitem [{\citenamefont {Adam}\ \emph
  {et~al.}(2016{\natexlab{a}})\citenamefont {Adam} \emph
  {et~al.}}]{Adam:2015lda}%
  \BibitemOpen
  \bibfield  {author} {\bibinfo {author} {\bibfnamefont {J.}~\bibnamefont
  {Adam}} \emph {et~al.} (\bibinfo {collaboration} {ALICE}),\ }\href {\doibase
  10.1016/j.physletb.2016.01.020} {\bibfield  {journal} {\bibinfo  {journal}
  {Phys. Lett. B}\ }\textbf {\bibinfo {volume} {754}},\ \bibinfo {pages} {235}
  (\bibinfo {year} {2016}{\natexlab{a}})},\ \Eprint
  {http://arxiv.org/abs/1509.07324} {arXiv:1509.07324 [nucl-ex]} \BibitemShut
  {NoStop}%
\bibitem [{\citenamefont {Adam}\ \emph
  {et~al.}(2016{\natexlab{b}})\citenamefont {Adam} \emph
  {et~al.}}]{Adam:2015vsf}%
  \BibitemOpen
  \bibfield  {author} {\bibinfo {author} {\bibfnamefont {J.}~\bibnamefont
  {Adam}} \emph {et~al.} (\bibinfo {collaboration} {ALICE}),\ }\href {\doibase
  10.1016/j.physletb.2016.05.027} {\bibfield  {journal} {\bibinfo  {journal}
  {Phys. Lett. B}\ }\textbf {\bibinfo {volume} {758}},\ \bibinfo {pages} {389}
  (\bibinfo {year} {2016}{\natexlab{b}})},\ \Eprint
  {http://arxiv.org/abs/1512.07227} {arXiv:1512.07227 [nucl-ex]} \BibitemShut
  {NoStop}%
\bibitem [{\citenamefont {Adam}\ \emph {et~al.}(2017)\citenamefont {Adam} \emph
  {et~al.}}]{ALICE:2017jyt}%
  \BibitemOpen
  \bibfield  {author} {\bibinfo {author} {\bibfnamefont {J.}~\bibnamefont
  {Adam}} \emph {et~al.} (\bibinfo {collaboration} {ALICE}),\ }\href {\doibase
  10.1038/nphys4111} {\bibfield  {journal} {\bibinfo  {journal} {Nature Phys.}\
  }\textbf {\bibinfo {volume} {13}},\ \bibinfo {pages} {535} (\bibinfo {year}
  {2017})},\ \Eprint {http://arxiv.org/abs/1606.07424} {arXiv:1606.07424
  [nucl-ex]} \BibitemShut {NoStop}%
\bibitem [{\citenamefont {Khachatryan}\ \emph {et~al.}(2015)\citenamefont
  {Khachatryan} \emph {et~al.}}]{Khachatryan:2015waa}%
  \BibitemOpen
  \bibfield  {author} {\bibinfo {author} {\bibfnamefont {V.}~\bibnamefont
  {Khachatryan}} \emph {et~al.} (\bibinfo {collaboration} {CMS}),\ }\href
  {\doibase 10.1103/PhysRevLett.115.012301} {\bibfield  {journal} {\bibinfo
  {journal} {Phys. Rev. Lett.}\ }\textbf {\bibinfo {volume} {115}},\ \bibinfo
  {pages} {012301} (\bibinfo {year} {2015})},\ \Eprint
  {http://arxiv.org/abs/1502.05382} {arXiv:1502.05382 [nucl-ex]} \BibitemShut
  {NoStop}%
\bibitem [{\citenamefont {Khachatryan}\ \emph
  {et~al.}(2017{\natexlab{a}})\citenamefont {Khachatryan} \emph
  {et~al.}}]{Khachatryan:2016txc}%
  \BibitemOpen
  \bibfield  {author} {\bibinfo {author} {\bibfnamefont {V.}~\bibnamefont
  {Khachatryan}} \emph {et~al.} (\bibinfo {collaboration} {CMS}),\ }\href
  {\doibase 10.1016/j.physletb.2016.12.009} {\bibfield  {journal} {\bibinfo
  {journal} {Phys. Lett. B}\ }\textbf {\bibinfo {volume} {765}},\ \bibinfo
  {pages} {193} (\bibinfo {year} {2017}{\natexlab{a}})},\ \Eprint
  {http://arxiv.org/abs/1606.06198} {arXiv:1606.06198 [nucl-ex]} \BibitemShut
  {NoStop}%
\bibitem [{\citenamefont {Citron}\ \emph {et~al.}(2019)\citenamefont {Citron}
  \emph {et~al.}}]{Citron:2018lsq}%
  \BibitemOpen
  \bibfield  {author} {\bibinfo {author} {\bibfnamefont {Z.}~\bibnamefont
  {Citron}} \emph {et~al.},\ }in\ \href {\doibase 10.23731/CYRM-2019-007.1159}
  {\emph {\bibinfo {booktitle} {Report on the Physics at the HL-LHC,and
  Perspectives for the HE-LHC}}},\ \bibinfo {editor} {edited by\ \bibinfo
  {editor} {\bibfnamefont {A.}~\bibnamefont {Dainese}}, \bibinfo {editor}
  {\bibfnamefont {M.}~\bibnamefont {Mangano}}, \bibinfo {editor} {\bibfnamefont
  {A.~B.}\ \bibnamefont {Meyer}}, \bibinfo {editor} {\bibfnamefont
  {A.}~\bibnamefont {Nisati}}, \bibinfo {editor} {\bibfnamefont
  {G.}~\bibnamefont {Salam}}, \ and\ \bibinfo {editor} {\bibfnamefont {M.~A.}\
  \bibnamefont {Vesterinen}}}\ (\bibinfo {year} {2019})\ pp.\ \bibinfo {pages}
  {1159--1410},\ \Eprint {http://arxiv.org/abs/1812.06772} {arXiv:1812.06772
  [hep-ph]} \BibitemShut {NoStop}%
\bibitem [{\citenamefont {Adolfsson}\ \emph {et~al.}(2020)\citenamefont
  {Adolfsson} \emph {et~al.}}]{Adolfsson:2020dhm}%
  \BibitemOpen
  \bibfield  {author} {\bibinfo {author} {\bibfnamefont {J.}~\bibnamefont
  {Adolfsson}} \emph {et~al.}\ }(\bibinfo {year} {2020})\ \Eprint
  {http://arxiv.org/abs/2003.10997} {arXiv:2003.10997 [hep-ph]} \BibitemShut
  {NoStop}%
\bibitem [{\citenamefont {Khachatryan}\ \emph
  {et~al.}(2017{\natexlab{b}})\citenamefont {Khachatryan} \emph
  {et~al.}}]{Khachatryan:2016odn}%
  \BibitemOpen
  \bibfield  {author} {\bibinfo {author} {\bibfnamefont {V.}~\bibnamefont
  {Khachatryan}} \emph {et~al.} (\bibinfo {collaboration} {CMS}),\ }\href
  {\doibase 10.1007/JHEP04(2017)039} {\bibfield  {journal} {\bibinfo  {journal}
  {JHEP}\ }\textbf {\bibinfo {volume} {04}},\ \bibinfo {pages} {039} (\bibinfo
  {year} {2017}{\natexlab{b}})},\ \Eprint {http://arxiv.org/abs/1611.01664}
  {arXiv:1611.01664 [nucl-ex]} \BibitemShut {NoStop}%
\bibitem [{\citenamefont {Sirunyan}\ \emph
  {et~al.}(2018{\natexlab{a}})\citenamefont {Sirunyan} \emph
  {et~al.}}]{Sirunyan:2018eqi}%
  \BibitemOpen
  \bibfield  {author} {\bibinfo {author} {\bibfnamefont {A.~M.}\ \bibnamefont
  {Sirunyan}} \emph {et~al.} (\bibinfo {collaboration} {CMS}),\ }\href
  {\doibase 10.1007/JHEP10(2018)138} {\bibfield  {journal} {\bibinfo  {journal}
  {JHEP}\ }\textbf {\bibinfo {volume} {10}},\ \bibinfo {pages} {138} (\bibinfo
  {year} {2018}{\natexlab{a}})},\ \Eprint {http://arxiv.org/abs/1809.00201}
  {arXiv:1809.00201 [hep-ex]} \BibitemShut {NoStop}%
\bibitem [{\citenamefont {Aaboud}\ \emph {et~al.}(2019)\citenamefont {Aaboud}
  \emph {et~al.}}]{Aaboud:2018twu}%
  \BibitemOpen
  \bibfield  {author} {\bibinfo {author} {\bibfnamefont {M.}~\bibnamefont
  {Aaboud}} \emph {et~al.} (\bibinfo {collaboration} {ATLAS}),\ }\href
  {\doibase 10.1016/j.physletb.2018.10.076} {\bibfield  {journal} {\bibinfo
  {journal} {Phys. Lett.}\ }\textbf {\bibinfo {volume} {B790}},\ \bibinfo
  {pages} {108} (\bibinfo {year} {2019})},\ \Eprint
  {http://arxiv.org/abs/1805.05635} {arXiv:1805.05635 [nucl-ex]} \BibitemShut
  {NoStop}%
\bibitem [{\citenamefont {Aad}\ \emph {et~al.}(2015{\natexlab{b}})\citenamefont
  {Aad} \emph {et~al.}}]{ATLAS:2014cpa}%
  \BibitemOpen
  \bibfield  {author} {\bibinfo {author} {\bibfnamefont {G.}~\bibnamefont
  {Aad}} \emph {et~al.} (\bibinfo {collaboration} {ATLAS}),\ }\href {\doibase
  10.1016/j.physletb.2015.07.023} {\bibfield  {journal} {\bibinfo  {journal}
  {Phys. Lett.}\ }\textbf {\bibinfo {volume} {B748}},\ \bibinfo {pages} {392}
  (\bibinfo {year} {2015}{\natexlab{b}})},\ \Eprint
  {http://arxiv.org/abs/1412.4092} {arXiv:1412.4092 [hep-ex]} \BibitemShut
  {NoStop}%
\bibitem [{\citenamefont {Glauber}\ and\ \citenamefont
  {Matthiae}(1970)}]{Glauber:1970jm}%
  \BibitemOpen
  \bibfield  {author} {\bibinfo {author} {\bibfnamefont {R.}~\bibnamefont
  {Glauber}}\ and\ \bibinfo {author} {\bibfnamefont {G.}~\bibnamefont
  {Matthiae}},\ }\href {\doibase 10.1016/0550-3213(70)90511-0} {\bibfield
  {journal} {\bibinfo  {journal} {Nucl. Phys. B}\ }\textbf {\bibinfo {volume}
  {21}},\ \bibinfo {pages} {135} (\bibinfo {year} {1970})}\BibitemShut
  {NoStop}%
\bibitem [{\citenamefont {d'Enterria}(2003)}]{dEnterria:2003xac}%
  \BibitemOpen
  \bibfield  {author} {\bibinfo {author} {\bibfnamefont {D.~G.}\ \bibnamefont
  {d'Enterria}},\ }\href@noop {} {\  (\bibinfo {year} {2003})},\ \Eprint
  {http://arxiv.org/abs/nucl-ex/0302016} {arXiv:nucl-ex/0302016} \BibitemShut
  {NoStop}%
\bibitem [{\citenamefont {Miller}\ \emph {et~al.}(2007)\citenamefont {Miller},
  \citenamefont {Reygers}, \citenamefont {Sanders},\ and\ \citenamefont
  {Steinberg}}]{Miller:2007ri}%
  \BibitemOpen
  \bibfield  {author} {\bibinfo {author} {\bibfnamefont {M.~L.}\ \bibnamefont
  {Miller}}, \bibinfo {author} {\bibfnamefont {K.}~\bibnamefont {Reygers}},
  \bibinfo {author} {\bibfnamefont {S.~J.}\ \bibnamefont {Sanders}}, \ and\
  \bibinfo {author} {\bibfnamefont {P.}~\bibnamefont {Steinberg}},\ }\href
  {\doibase 10.1146/annurev.nucl.57.090506.123020} {\bibfield  {journal}
  {\bibinfo  {journal} {Ann. Rev. Nucl. Part. Sci.}\ }\textbf {\bibinfo
  {volume} {57}},\ \bibinfo {pages} {205} (\bibinfo {year} {2007})},\ \Eprint
  {http://arxiv.org/abs/nucl-ex/0701025} {arXiv:nucl-ex/0701025} \BibitemShut
  {NoStop}%
\bibitem [{\citenamefont {Loizides}\ and\ \citenamefont
  {Morsch}(2017)}]{Morsch:2017brb}%
  \BibitemOpen
  \bibfield  {author} {\bibinfo {author} {\bibfnamefont {C.}~\bibnamefont
  {Loizides}}\ and\ \bibinfo {author} {\bibfnamefont {A.}~\bibnamefont
  {Morsch}},\ }\href {\doibase 10.1016/j.physletb.2017.09.002} {\bibfield
  {journal} {\bibinfo  {journal} {Phys. Lett.}\ }\textbf {\bibinfo {volume}
  {B773}},\ \bibinfo {pages} {408} (\bibinfo {year} {2017})},\ \Eprint
  {http://arxiv.org/abs/1705.08856} {arXiv:1705.08856 [nucl-ex]} \BibitemShut
  {NoStop}%
\bibitem [{\citenamefont {Eskola}\ \emph {et~al.}(2020)\citenamefont {Eskola},
  \citenamefont {Helenius}, \citenamefont {Kuha},\ and\ \citenamefont
  {Paukkunen}}]{Eskola:2020lee}%
  \BibitemOpen
  \bibfield  {author} {\bibinfo {author} {\bibfnamefont {K.~J.}\ \bibnamefont
  {Eskola}}, \bibinfo {author} {\bibfnamefont {I.}~\bibnamefont {Helenius}},
  \bibinfo {author} {\bibfnamefont {M.}~\bibnamefont {Kuha}}, \ and\ \bibinfo
  {author} {\bibfnamefont {H.}~\bibnamefont {Paukkunen}},\ }\href@noop {} {\
  (\bibinfo {year} {2020})},\ \Eprint {http://arxiv.org/abs/2003.11856}
  {arXiv:2003.11856 [hep-ph]} \BibitemShut {NoStop}%
\bibitem [{\citenamefont {Currie}\ \emph {et~al.}(2017)\citenamefont {Currie},
  \citenamefont {Glover},\ and\ \citenamefont {Pires}}]{Currie:2016bfm}%
  \BibitemOpen
  \bibfield  {author} {\bibinfo {author} {\bibfnamefont {J.}~\bibnamefont
  {Currie}}, \bibinfo {author} {\bibfnamefont {E.}~\bibnamefont {Glover}}, \
  and\ \bibinfo {author} {\bibfnamefont {J.}~\bibnamefont {Pires}},\ }\href
  {\doibase 10.1103/PhysRevLett.118.072002} {\bibfield  {journal} {\bibinfo
  {journal} {Phys. Rev. Lett.}\ }\textbf {\bibinfo {volume} {118}},\ \bibinfo
  {pages} {072002} (\bibinfo {year} {2017})},\ \Eprint
  {http://arxiv.org/abs/1611.01460} {arXiv:1611.01460 [hep-ph]} \BibitemShut
  {NoStop}%
\bibitem [{\citenamefont {Gehrmann}\ \emph {et~al.}(2018)\citenamefont
  {Gehrmann} \emph {et~al.}}]{Gehrmann:2018szu}%
  \BibitemOpen
  \bibfield  {author} {\bibinfo {author} {\bibfnamefont {T.}~\bibnamefont
  {Gehrmann}} \emph {et~al.},\ }\href {\doibase 10.22323/1.290.0074} {\bibfield
   {journal} {\bibinfo  {journal} {PoS}\ }\textbf {\bibinfo {volume}
  {RADCOR2017}},\ \bibinfo {pages} {074} (\bibinfo {year} {2018})},\ \Eprint
  {http://arxiv.org/abs/1801.06415} {arXiv:1801.06415 [hep-ph]} \BibitemShut
  {NoStop}%
\bibitem [{\citenamefont {Britzger}\ \emph {et~al.}(2019)\citenamefont
  {Britzger} \emph {et~al.}}]{Britzger:2019kkb}%
  \BibitemOpen
  \bibfield  {author} {\bibinfo {author} {\bibfnamefont {D.}~\bibnamefont
  {Britzger}} \emph {et~al.},\ }\href {\doibase 10.1140/epjc/s10052-019-7351-x}
  {\bibfield  {journal} {\bibinfo  {journal} {Eur. Phys. J. C}\ }\textbf
  {\bibinfo {volume} {79}},\ \bibinfo {pages} {845} (\bibinfo {year} {2019})},\
  \Eprint {http://arxiv.org/abs/1906.05303} {arXiv:1906.05303 [hep-ph]}
  \BibitemShut {NoStop}%
\bibitem [{\citenamefont {Dulat}\ \emph {et~al.}(2016)\citenamefont {Dulat},
  \citenamefont {Hou}, \citenamefont {Gao}, \citenamefont {Guzzi},
  \citenamefont {Huston}, \citenamefont {Nadolsky}, \citenamefont {Pumplin},
  \citenamefont {Schmidt}, \citenamefont {Stump},\ and\ \citenamefont
  {Yuan}}]{Dulat:2015mca}%
  \BibitemOpen
  \bibfield  {author} {\bibinfo {author} {\bibfnamefont {S.}~\bibnamefont
  {Dulat}}, \bibinfo {author} {\bibfnamefont {T.-J.}\ \bibnamefont {Hou}},
  \bibinfo {author} {\bibfnamefont {J.}~\bibnamefont {Gao}}, \bibinfo {author}
  {\bibfnamefont {M.}~\bibnamefont {Guzzi}}, \bibinfo {author} {\bibfnamefont
  {J.}~\bibnamefont {Huston}}, \bibinfo {author} {\bibfnamefont
  {P.}~\bibnamefont {Nadolsky}}, \bibinfo {author} {\bibfnamefont
  {J.}~\bibnamefont {Pumplin}}, \bibinfo {author} {\bibfnamefont
  {C.}~\bibnamefont {Schmidt}}, \bibinfo {author} {\bibfnamefont
  {D.}~\bibnamefont {Stump}}, \ and\ \bibinfo {author} {\bibfnamefont
  {C.}~\bibnamefont {Yuan}},\ }\href {\doibase 10.1103/PhysRevD.93.033006}
  {\bibfield  {journal} {\bibinfo  {journal} {Phys. Rev. D}\ }\textbf {\bibinfo
  {volume} {93}},\ \bibinfo {pages} {033006} (\bibinfo {year} {2016})},\
  \Eprint {http://arxiv.org/abs/1506.07443} {arXiv:1506.07443 [hep-ph]}
  \BibitemShut {NoStop}%
\bibitem [{\citenamefont {Alioli}\ \emph {et~al.}(2011)\citenamefont {Alioli},
  \citenamefont {Hamilton}, \citenamefont {Nason}, \citenamefont {Oleari},\
  and\ \citenamefont {Re}}]{Alioli:2010xa}%
  \BibitemOpen
  \bibfield  {author} {\bibinfo {author} {\bibfnamefont {S.}~\bibnamefont
  {Alioli}}, \bibinfo {author} {\bibfnamefont {K.}~\bibnamefont {Hamilton}},
  \bibinfo {author} {\bibfnamefont {P.}~\bibnamefont {Nason}}, \bibinfo
  {author} {\bibfnamefont {C.}~\bibnamefont {Oleari}}, \ and\ \bibinfo {author}
  {\bibfnamefont {E.}~\bibnamefont {Re}},\ }\href {\doibase
  10.1007/JHEP04(2011)081} {\bibfield  {journal} {\bibinfo  {journal} {JHEP}\
  }\textbf {\bibinfo {volume} {04}},\ \bibinfo {pages} {081} (\bibinfo {year}
  {2011})},\ \Eprint {http://arxiv.org/abs/1012.3380} {arXiv:1012.3380
  [hep-ph]} \BibitemShut {NoStop}%
\bibitem [{\citenamefont {Eskola}\ \emph {et~al.}(2017)\citenamefont {Eskola},
  \citenamefont {Paakkinen}, \citenamefont {Paukkunen},\ and\ \citenamefont
  {Salgado}}]{Eskola:2016oht}%
  \BibitemOpen
  \bibfield  {author} {\bibinfo {author} {\bibfnamefont {K.~J.}\ \bibnamefont
  {Eskola}}, \bibinfo {author} {\bibfnamefont {P.}~\bibnamefont {Paakkinen}},
  \bibinfo {author} {\bibfnamefont {H.}~\bibnamefont {Paukkunen}}, \ and\
  \bibinfo {author} {\bibfnamefont {C.~A.}\ \bibnamefont {Salgado}},\ }\href
  {\doibase 10.1140/epjc/s10052-017-4725-9} {\bibfield  {journal} {\bibinfo
  {journal} {Eur. Phys. J.}\ }\textbf {\bibinfo {volume} {C77}},\ \bibinfo
  {pages} {163} (\bibinfo {year} {2017})},\ \Eprint
  {http://arxiv.org/abs/1612.05741} {arXiv:1612.05741 [hep-ph]} \BibitemShut
  {NoStop}%
\bibitem [{\citenamefont {Sirunyan}\ \emph
  {et~al.}(2018{\natexlab{b}})\citenamefont {Sirunyan} \emph
  {et~al.}}]{Sirunyan:2018qel}%
  \BibitemOpen
  \bibfield  {author} {\bibinfo {author} {\bibfnamefont {A.~M.}\ \bibnamefont
  {Sirunyan}} \emph {et~al.} (\bibinfo {collaboration} {CMS}),\ }\href
  {\doibase 10.1103/PhysRevLett.121.062002} {\bibfield  {journal} {\bibinfo
  {journal} {Phys. Rev. Lett.}\ }\textbf {\bibinfo {volume} {121}},\ \bibinfo
  {pages} {062002} (\bibinfo {year} {2018}{\natexlab{b}})},\ \Eprint
  {http://arxiv.org/abs/1805.04736} {arXiv:1805.04736 [hep-ex]} \BibitemShut
  {NoStop}%
\bibitem [{\citenamefont {Paukkunen}\ and\ \citenamefont
  {Zurita}(2014)}]{Paukkunen:2014zia}%
  \BibitemOpen
  \bibfield  {author} {\bibinfo {author} {\bibfnamefont {H.}~\bibnamefont
  {Paukkunen}}\ and\ \bibinfo {author} {\bibfnamefont {P.}~\bibnamefont
  {Zurita}},\ }\href {\doibase 10.1007/JHEP12(2014)100} {\bibfield  {journal}
  {\bibinfo  {journal} {JHEP}\ }\textbf {\bibinfo {volume} {12}},\ \bibinfo
  {pages} {100} (\bibinfo {year} {2014})},\ \Eprint
  {http://arxiv.org/abs/1402.6623} {arXiv:1402.6623 [hep-ph]} \BibitemShut
  {NoStop}%
\bibitem [{\citenamefont {Eskola}\ \emph {et~al.}(2019)\citenamefont {Eskola},
  \citenamefont {Paakkinen},\ and\ \citenamefont {Paukkunen}}]{Eskola:2019dui}%
  \BibitemOpen
  \bibfield  {author} {\bibinfo {author} {\bibfnamefont {K.~J.}\ \bibnamefont
  {Eskola}}, \bibinfo {author} {\bibfnamefont {P.}~\bibnamefont {Paakkinen}}, \
  and\ \bibinfo {author} {\bibfnamefont {H.}~\bibnamefont {Paukkunen}},\ }\href
  {\doibase 10.1140/epjc/s10052-019-6982-2} {\bibfield  {journal} {\bibinfo
  {journal} {Eur. Phys. J. C}\ }\textbf {\bibinfo {volume} {79}},\ \bibinfo
  {pages} {511} (\bibinfo {year} {2019})},\ \Eprint
  {http://arxiv.org/abs/1903.09832} {arXiv:1903.09832 [hep-ph]} \BibitemShut
  {NoStop}%
\bibitem [{ATL(2019)}]{ATLAS-CONF-2019-021}%
  \BibitemOpen
  \href {http://cds.cern.ch/record/2677054} {\emph {\bibinfo {title}
  {{Luminosity determination in $pp$ collisions at $\sqrt{s}=13$ TeV using the
  ATLAS detector at the LHC}}}},\ \bibinfo {type} {Tech. Rep.}\ \bibinfo
  {number} {ATLAS-CONF-2019-021}\ (\bibinfo  {institution} {CERN},\ \bibinfo
  {address} {Geneva},\ \bibinfo {year} {2019})\BibitemShut {NoStop}%
\bibitem [{ATL(2020)}]{ATLAS-CONF-2020-010}%
  \BibitemOpen
  \href {http://cds.cern.ch/record/2719516} {\emph {\bibinfo {title}
  {{Measurement of light-by-light scattering and search for axion-like
  particles with 2.2 nb$^{-1}$ of Pb+Pb data with the ATLAS detector}}}},\
  \bibinfo {type} {Tech. Rep.}\ \bibinfo {number} {ATLAS-CONF-2020-010}\
  (\bibinfo  {institution} {CERN},\ \bibinfo {address} {Geneva},\ \bibinfo
  {year} {2020})\BibitemShut {NoStop}%
\bibitem [{\citenamefont {Binnewies}\ \emph {et~al.}(1995)\citenamefont
  {Binnewies}, \citenamefont {Kniehl},\ and\ \citenamefont
  {Kramer}}]{Binnewies:1994ju}%
  \BibitemOpen
  \bibfield  {author} {\bibinfo {author} {\bibfnamefont {J.}~\bibnamefont
  {Binnewies}}, \bibinfo {author} {\bibfnamefont {B.~A.}\ \bibnamefont
  {Kniehl}}, \ and\ \bibinfo {author} {\bibfnamefont {G.}~\bibnamefont
  {Kramer}},\ }\href {\doibase 10.1007/BF01556135} {\bibfield  {journal}
  {\bibinfo  {journal} {Z. Phys. C}\ }\textbf {\bibinfo {volume} {65}},\
  \bibinfo {pages} {471} (\bibinfo {year} {1995})},\ \Eprint
  {http://arxiv.org/abs/hep-ph/9407347} {arXiv:hep-ph/9407347} \BibitemShut
  {NoStop}%
\bibitem [{\citenamefont {Kniehl}\ \emph {et~al.}(2000)\citenamefont {Kniehl},
  \citenamefont {Kramer},\ and\ \citenamefont {Potter}}]{Kniehl:2000fe}%
  \BibitemOpen
  \bibfield  {author} {\bibinfo {author} {\bibfnamefont {B.~A.}\ \bibnamefont
  {Kniehl}}, \bibinfo {author} {\bibfnamefont {G.}~\bibnamefont {Kramer}}, \
  and\ \bibinfo {author} {\bibfnamefont {B.}~\bibnamefont {Potter}},\ }\href
  {\doibase 10.1016/S0550-3213(00)00303-5} {\bibfield  {journal} {\bibinfo
  {journal} {Nucl. Phys.}\ }\textbf {\bibinfo {volume} {B582}},\ \bibinfo
  {pages} {514} (\bibinfo {year} {2000})},\ \Eprint
  {http://arxiv.org/abs/hep-ph/0010289} {arXiv:hep-ph/0010289 [hep-ph]}
  \BibitemShut {NoStop}%
\bibitem [{\citenamefont {Aversa}\ \emph {et~al.}(1989)\citenamefont {Aversa},
  \citenamefont {Chiappetta}, \citenamefont {Greco},\ and\ \citenamefont
  {Guillet}}]{Aversa:1988vb}%
  \BibitemOpen
  \bibfield  {author} {\bibinfo {author} {\bibfnamefont {F.}~\bibnamefont
  {Aversa}}, \bibinfo {author} {\bibfnamefont {P.}~\bibnamefont {Chiappetta}},
  \bibinfo {author} {\bibfnamefont {M.}~\bibnamefont {Greco}}, \ and\ \bibinfo
  {author} {\bibfnamefont {J.}~\bibnamefont {Guillet}},\ }\href {\doibase
  10.1016/0550-3213(89)90288-5} {\bibfield  {journal} {\bibinfo  {journal}
  {Nucl. Phys. B}\ }\textbf {\bibinfo {volume} {327}},\ \bibinfo {pages} {105}
  (\bibinfo {year} {1989})}\BibitemShut {NoStop}%
\bibitem [{Note1()}]{Note1}%
  \BibitemOpen
  \bibinfo {note} {\protect \url
  {http://lapth.cnrs.fr/PHOX_FAMILY/readme_inc.html}}\BibitemShut {NoStop}%
\bibitem [{\citenamefont {Buckley}\ \emph {et~al.}(2015)\citenamefont
  {Buckley}, \citenamefont {Ferrando}, \citenamefont {Lloyd}, \citenamefont
  {Nordström}, \citenamefont {Page}, \citenamefont {Rüfenacht}, \citenamefont
  {Schönherr},\ and\ \citenamefont {Watt}}]{Buckley:2014ana}%
  \BibitemOpen
  \bibfield  {author} {\bibinfo {author} {\bibfnamefont {A.}~\bibnamefont
  {Buckley}}, \bibinfo {author} {\bibfnamefont {J.}~\bibnamefont {Ferrando}},
  \bibinfo {author} {\bibfnamefont {S.}~\bibnamefont {Lloyd}}, \bibinfo
  {author} {\bibfnamefont {K.}~\bibnamefont {Nordström}}, \bibinfo {author}
  {\bibfnamefont {B.}~\bibnamefont {Page}}, \bibinfo {author} {\bibfnamefont
  {M.}~\bibnamefont {Rüfenacht}}, \bibinfo {author} {\bibfnamefont
  {M.}~\bibnamefont {Schönherr}}, \ and\ \bibinfo {author} {\bibfnamefont
  {G.}~\bibnamefont {Watt}},\ }\href {\doibase 10.1140/epjc/s10052-015-3318-8}
  {\bibfield  {journal} {\bibinfo  {journal} {Eur. Phys. J.}\ }\textbf
  {\bibinfo {volume} {C75}},\ \bibinfo {pages} {132} (\bibinfo {year}
  {2015})},\ \Eprint {http://arxiv.org/abs/1412.7420} {arXiv:1412.7420
  [hep-ph]} \BibitemShut {NoStop}%
\bibitem [{\citenamefont {Huss}\ \emph {et~al.}(2020)\citenamefont {Huss},
  \citenamefont {Kurkela}, \citenamefont {Mazeliauskas}, \citenamefont
  {Paatelainen}, \citenamefont {van~der Schee},\ and\ \citenamefont
  {Wiedemann}}]{Huss:2020whe}%
  \BibitemOpen
  \bibfield  {author} {\bibinfo {author} {\bibfnamefont {A.}~\bibnamefont
  {Huss}}, \bibinfo {author} {\bibfnamefont {A.}~\bibnamefont {Kurkela}},
  \bibinfo {author} {\bibfnamefont {A.}~\bibnamefont {Mazeliauskas}}, \bibinfo
  {author} {\bibfnamefont {R.}~\bibnamefont {Paatelainen}}, \bibinfo {author}
  {\bibfnamefont {W.}~\bibnamefont {van~der Schee}}, \ and\ \bibinfo {author}
  {\bibfnamefont {U.~A.}\ \bibnamefont {Wiedemann}},\ }\href@noop {} {\
  (\bibinfo {year} {2020})},\ \Eprint {http://arxiv.org/abs/2007.13758}
  {arXiv:2007.13758 [hep-ph]} \BibitemShut {NoStop}%
\bibitem [{\citenamefont {Zapp}(2014)}]{Zapp:2013vla}%
  \BibitemOpen
  \bibfield  {author} {\bibinfo {author} {\bibfnamefont {K.~C.}\ \bibnamefont
  {Zapp}},\ }\href {\doibase 10.1140/epjc/s10052-014-2762-1} {\bibfield
  {journal} {\bibinfo  {journal} {Eur. Phys. J. C}\ }\textbf {\bibinfo {volume}
  {74}},\ \bibinfo {pages} {2762} (\bibinfo {year} {2014})},\ \Eprint
  {http://arxiv.org/abs/1311.0048} {arXiv:1311.0048 [hep-ph]} \BibitemShut
  {NoStop}%
\bibitem [{\citenamefont {Putschke}\ \emph {et~al.}(2019)\citenamefont
  {Putschke} \emph {et~al.}}]{Putschke:2019yrg}%
  \BibitemOpen
  \bibfield  {author} {\bibinfo {author} {\bibfnamefont {J.}~\bibnamefont
  {Putschke}} \emph {et~al.},\ }\href@noop {} {\  (\bibinfo {year} {2019})},\
  \Eprint {http://arxiv.org/abs/1903.07706} {arXiv:1903.07706 [nucl-th]}
  \BibitemShut {NoStop}%
\bibitem [{\citenamefont {Schenke}\ \emph {et~al.}(2009)\citenamefont
  {Schenke}, \citenamefont {Gale},\ and\ \citenamefont
  {Jeon}}]{Schenke:2009gb}%
  \BibitemOpen
  \bibfield  {author} {\bibinfo {author} {\bibfnamefont {B.}~\bibnamefont
  {Schenke}}, \bibinfo {author} {\bibfnamefont {C.}~\bibnamefont {Gale}}, \
  and\ \bibinfo {author} {\bibfnamefont {S.}~\bibnamefont {Jeon}},\ }\href
  {\doibase 10.1103/PhysRevC.80.054913} {\bibfield  {journal} {\bibinfo
  {journal} {Phys. Rev. C}\ }\textbf {\bibinfo {volume} {80}},\ \bibinfo
  {pages} {054913} (\bibinfo {year} {2009})},\ \Eprint
  {http://arxiv.org/abs/0909.2037} {arXiv:0909.2037 [hep-ph]} \BibitemShut
  {NoStop}%
\bibitem [{\citenamefont {Baier}\ \emph
  {et~al.}(1997{\natexlab{a}})\citenamefont {Baier}, \citenamefont
  {Dokshitzer}, \citenamefont {Mueller}, \citenamefont {Peigne},\ and\
  \citenamefont {Schiff}}]{Baier:1996kr}%
  \BibitemOpen
  \bibfield  {author} {\bibinfo {author} {\bibfnamefont {R.}~\bibnamefont
  {Baier}}, \bibinfo {author} {\bibfnamefont {Y.~L.}\ \bibnamefont
  {Dokshitzer}}, \bibinfo {author} {\bibfnamefont {A.~H.}\ \bibnamefont
  {Mueller}}, \bibinfo {author} {\bibfnamefont {S.}~\bibnamefont {Peigne}}, \
  and\ \bibinfo {author} {\bibfnamefont {D.}~\bibnamefont {Schiff}},\ }\href
  {\doibase 10.1016/S0550-3213(96)00553-6} {\bibfield  {journal} {\bibinfo
  {journal} {Nucl. Phys. B}\ }\textbf {\bibinfo {volume} {483}},\ \bibinfo
  {pages} {291} (\bibinfo {year} {1997}{\natexlab{a}})},\ \Eprint
  {http://arxiv.org/abs/hep-ph/9607355} {arXiv:hep-ph/9607355} \BibitemShut
  {NoStop}%
\bibitem [{\citenamefont {Baier}\ \emph
  {et~al.}(1997{\natexlab{b}})\citenamefont {Baier}, \citenamefont
  {Dokshitzer}, \citenamefont {Mueller}, \citenamefont {Peigne},\ and\
  \citenamefont {Schiff}}]{Baier:1996sk}%
  \BibitemOpen
  \bibfield  {author} {\bibinfo {author} {\bibfnamefont {R.}~\bibnamefont
  {Baier}}, \bibinfo {author} {\bibfnamefont {Y.~L.}\ \bibnamefont
  {Dokshitzer}}, \bibinfo {author} {\bibfnamefont {A.~H.}\ \bibnamefont
  {Mueller}}, \bibinfo {author} {\bibfnamefont {S.}~\bibnamefont {Peigne}}, \
  and\ \bibinfo {author} {\bibfnamefont {D.}~\bibnamefont {Schiff}},\ }\href
  {\doibase 10.1016/S0550-3213(96)00581-0} {\bibfield  {journal} {\bibinfo
  {journal} {Nucl. Phys. B}\ }\textbf {\bibinfo {volume} {484}},\ \bibinfo
  {pages} {265} (\bibinfo {year} {1997}{\natexlab{b}})},\ \Eprint
  {http://arxiv.org/abs/hep-ph/9608322} {arXiv:hep-ph/9608322} \BibitemShut
  {NoStop}%
\bibitem [{\citenamefont {Zakharov}(1996)}]{Zakharov:1996fv}%
  \BibitemOpen
  \bibfield  {author} {\bibinfo {author} {\bibfnamefont {B.}~\bibnamefont
  {Zakharov}},\ }\href {\doibase 10.1134/1.567126} {\bibfield  {journal}
  {\bibinfo  {journal} {JETP Lett.}\ }\textbf {\bibinfo {volume} {63}},\
  \bibinfo {pages} {952} (\bibinfo {year} {1996})},\ \Eprint
  {http://arxiv.org/abs/hep-ph/9607440} {arXiv:hep-ph/9607440} \BibitemShut
  {NoStop}%
\bibitem [{\citenamefont {Zakharov}(1997)}]{Zakharov:1997uu}%
  \BibitemOpen
  \bibfield  {author} {\bibinfo {author} {\bibfnamefont {B.}~\bibnamefont
  {Zakharov}},\ }\href {\doibase 10.1134/1.567389} {\bibfield  {journal}
  {\bibinfo  {journal} {JETP Lett.}\ }\textbf {\bibinfo {volume} {65}},\
  \bibinfo {pages} {615} (\bibinfo {year} {1997})},\ \Eprint
  {http://arxiv.org/abs/hep-ph/9704255} {arXiv:hep-ph/9704255} \BibitemShut
  {NoStop}%
\bibitem [{\citenamefont {Wiedemann}(2000)}]{Wiedemann:2000za}%
  \BibitemOpen
  \bibfield  {author} {\bibinfo {author} {\bibfnamefont {U.~A.}\ \bibnamefont
  {Wiedemann}},\ }\href {\doibase 10.1016/S0550-3213(00)00457-0} {\bibfield
  {journal} {\bibinfo  {journal} {Nucl. Phys. B}\ }\textbf {\bibinfo {volume}
  {588}},\ \bibinfo {pages} {303} (\bibinfo {year} {2000})},\ \Eprint
  {http://arxiv.org/abs/hep-ph/0005129} {arXiv:hep-ph/0005129} \BibitemShut
  {NoStop}%
\bibitem [{\citenamefont {Gyulassy}\ \emph {et~al.}(2001)\citenamefont
  {Gyulassy}, \citenamefont {Levai},\ and\ \citenamefont
  {Vitev}}]{Gyulassy:2000er}%
  \BibitemOpen
  \bibfield  {author} {\bibinfo {author} {\bibfnamefont {M.}~\bibnamefont
  {Gyulassy}}, \bibinfo {author} {\bibfnamefont {P.}~\bibnamefont {Levai}}, \
  and\ \bibinfo {author} {\bibfnamefont {I.}~\bibnamefont {Vitev}},\ }\href
  {\doibase 10.1016/S0550-3213(00)00652-0} {\bibfield  {journal} {\bibinfo
  {journal} {Nucl. Phys. B}\ }\textbf {\bibinfo {volume} {594}},\ \bibinfo
  {pages} {371} (\bibinfo {year} {2001})},\ \Eprint
  {http://arxiv.org/abs/nucl-th/0006010} {arXiv:nucl-th/0006010} \BibitemShut
  {NoStop}%
\bibitem [{\citenamefont {Wang}\ and\ \citenamefont
  {Guo}(2001)}]{Wang:2001ifa}%
  \BibitemOpen
  \bibfield  {author} {\bibinfo {author} {\bibfnamefont {X.-N.}\ \bibnamefont
  {Wang}}\ and\ \bibinfo {author} {\bibfnamefont {X.-f.}\ \bibnamefont {Guo}},\
  }\href {\doibase 10.1016/S0375-9474(01)01130-7} {\bibfield  {journal}
  {\bibinfo  {journal} {Nucl. Phys. A}\ }\textbf {\bibinfo {volume} {696}},\
  \bibinfo {pages} {788} (\bibinfo {year} {2001})},\ \Eprint
  {http://arxiv.org/abs/hep-ph/0102230} {arXiv:hep-ph/0102230} \BibitemShut
  {NoStop}%
\bibitem [{\citenamefont {Armesto}\ \emph {et~al.}(2012)\citenamefont {Armesto}
  \emph {et~al.}}]{Armesto:2011ht}%
  \BibitemOpen
  \bibfield  {author} {\bibinfo {author} {\bibfnamefont {N.}~\bibnamefont
  {Armesto}} \emph {et~al.},\ }\href {\doibase 10.1103/PhysRevC.86.064904}
  {\bibfield  {journal} {\bibinfo  {journal} {Phys. Rev. C}\ }\textbf {\bibinfo
  {volume} {86}},\ \bibinfo {pages} {064904} (\bibinfo {year} {2012})},\
  \Eprint {http://arxiv.org/abs/1106.1106} {arXiv:1106.1106 [hep-ph]}
  \BibitemShut {NoStop}%
\bibitem [{\citenamefont {Dittmar}\ \emph {et~al.}(1997)\citenamefont
  {Dittmar}, \citenamefont {Pauss},\ and\ \citenamefont
  {Zurcher}}]{Dittmar:1997md}%
  \BibitemOpen
  \bibfield  {author} {\bibinfo {author} {\bibfnamefont {M.}~\bibnamefont
  {Dittmar}}, \bibinfo {author} {\bibfnamefont {F.}~\bibnamefont {Pauss}}, \
  and\ \bibinfo {author} {\bibfnamefont {D.}~\bibnamefont {Zurcher}},\ }\href
  {\doibase 10.1103/PhysRevD.56.7284} {\bibfield  {journal} {\bibinfo
  {journal} {Phys. Rev. D}\ }\textbf {\bibinfo {volume} {56}},\ \bibinfo
  {pages} {7284} (\bibinfo {year} {1997})},\ \Eprint
  {http://arxiv.org/abs/hep-ex/9705004} {arXiv:hep-ex/9705004} \BibitemShut
  {NoStop}%
\bibitem [{\citenamefont {Khoze}\ \emph {et~al.}(2001)\citenamefont {Khoze},
  \citenamefont {Martin}, \citenamefont {Orava},\ and\ \citenamefont
  {Ryskin}}]{Khoze:2000db}%
  \BibitemOpen
  \bibfield  {author} {\bibinfo {author} {\bibfnamefont {V.~A.}\ \bibnamefont
  {Khoze}}, \bibinfo {author} {\bibfnamefont {A.~D.}\ \bibnamefont {Martin}},
  \bibinfo {author} {\bibfnamefont {R.}~\bibnamefont {Orava}}, \ and\ \bibinfo
  {author} {\bibfnamefont {M.}~\bibnamefont {Ryskin}},\ }\href {\doibase
  10.1007/s100520100616} {\bibfield  {journal} {\bibinfo  {journal} {Eur. Phys.
  J. C}\ }\textbf {\bibinfo {volume} {19}},\ \bibinfo {pages} {313} (\bibinfo
  {year} {2001})},\ \Eprint {http://arxiv.org/abs/hep-ph/0010163}
  {arXiv:hep-ph/0010163} \BibitemShut {NoStop}%
\end{thebibliography}%

\fakesection{Reweighting of Hessian nPDF sets\label{sec:reweight}}
The process independent nuclear parton distribution functions are extracted from global fits to a wide range of experimental data~\cite{Eskola:2016oht}. 
The impact of additional experimental data on nPDFs with Hessian error sets can be assessed via a reweighting procedure~\cite{Paukkunen:2014zia}. In Ref.~\cite{Eskola:2019dui} it was shown that including the LHC dijet data of $p$Pb collisions significantly reduces the nPDF uncertainties for EPPS16 nPDF sets. We have independently reproduced this calculation to determine the reweighting effect on jet and $Z$-boson production in OO collisions.
In the following, we briefly summarize this procedure.

In the Hessian approach, the parton distribution functions $f(x,Q)$ are parametrized by $z_j$ ($j=1,\ldots N$) internal parameters and a fit is performed by determining the global minimum $\bm{z}_\text{min}$ of the $\chi^2(\bm{z})$-function. 
Further, the $2N$ error sets represent the $\pm$ displacement around the minimum along the eigendirections of the Hessian matrix. The allowed range in the displacement is determined by some suitably chosen tolerance $\Delta \chi^2$, e.g. $\Delta \chi^2=52$ for EPPS16~\cite{Eskola:2016oht}.
The theoretical uncertainty of a given
observable $y_a$ around the central prediction
$y_a[\bm{z}_\text{min}]$ is then given by 
\begin{equation}
    \Delta y_a = \sqrt{\sum_k D_{ak}^2},
\end{equation}
where $D_{ak}$ is the difference of the observable evaluated on the $\pm$ error sets, i.e.,
\begin{equation}
D_{ak} = \frac{y_a[\bm{z}^+_k]-y_a[\bm{z}^-_k]}{2}.
\end{equation}
Note that for the discussion of the reweighting we consider the symmetric nPDF errors~\cite{Eskola:2016oht}.

In order to assess the impact of new data points $y_a^\text{data}$ on the nPDFs, we consider the $\chi^2$ after the inclusion of the new data
\begin{equation}
\chi^2_\text{new}(\bm{z}) =\chi^2(\bm{z})+  (y_a[\bm{z}] - y_a^\text{data})C_{ab}^{-1}    (y_b[\bm{z}] - y_b^\text{data}),
\end{equation}
where $C_{ab}$ is the covariance matrix of the measurement and $y_a[\bm{z}]$ the theory predictions.
Close to the initial global minimum the original $\chi^2$ can be approximated by a quadratic function in deviations from the minimum, while $y_a[\bm{z}]$ is linearized using $D_{ak}$. The new data shifts the location of the minimum and changes the Hessian around it~\cite{Paukkunen:2014zia}. The reweighted theoretical uncertainties are given by
\begin{equation}
    \Delta y_a = \sqrt{\sum_{s}\frac{1}{\lambda_s} ( \sum_k D_{ak}v^s_k)^2},
    \label{eq:newerr}
\end{equation}
where $\lambda_s$ and $ v_k^s$ are the $s$-th eigenvalue and normalized eigenvector of the matrix
\begin{align}
B_{ks}=\delta_{ks}+\frac{1}{\Delta \chi^2} D_{ak}C_{ab}^{-1}  D_{bs}.
\end{align}
Here $v_k^s$ represents the rotation of the Hessian matrix eigendirections, while $\lambda_s$ quantifies the reduction of nPDF uncertainties in that direction.
The theory prediction at the new minimum is given by
\begin{align}
  y_a[\bm{z}_\text{min}^\text{new}] &= 
  y_a[\bm{z}_\text{min}] 
  \nonumber\\-&\frac{1}{\Delta \chi^2} D_{ak}B_{ks}^{-1} D_{bs}C_{bc}^{-1}  (y_c[\bm{z}_\text{min}]-y_c^\text{data}).\label{eq:newmin}
\end{align}

\begin{figure}
    \centering
\includegraphics[width=\linewidth]{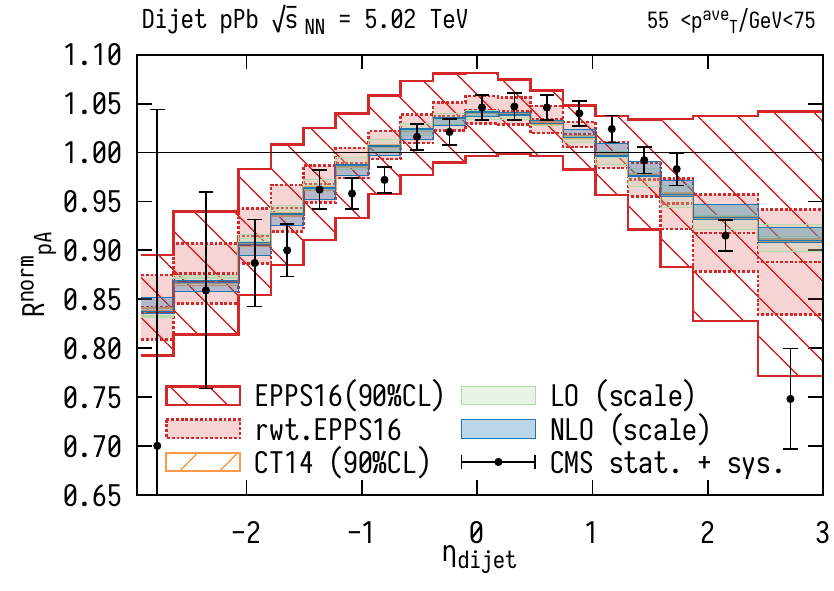}
    \caption{ Normalized dijet nuclear modification factor, \Eq{RAAdijet}.
     The open  red band shows initial nPDF uncertainties and the solid band---after reweighting
     in \Eqs{eq:newerr} and \eq{eq:newmin}. 
  The orange and blue bands show cancellation of fully correlated proton PDF and scale uncertainties.
    The error bars are combined experimental statistical and systematic uncertainties~\cite{Sirunyan:2018qel}. C.f. Fig.\,10 in Ref.~\cite{Eskola:2019dui}}
    \label{fig5}
\end{figure}

We follow the analysis of Ref.~\cite{Eskola:2019dui} and apply a reweighting to CMS $\sqrt{s_{NN}}=5.02\,\text{TeV}$  $p$Pb dijet data to quadratic order (beyond-quadratic terms were found not to be important for this data set).
Specifically, we consider the normalized dijet spectra ratio in some $p_T$ range
\begin{equation}
    R_{p\text{Pb}}^\text{norm} =\frac{1}{d\sigma^{p\text{Pb}}/dp_T}\frac{d\sigma^{p\text{Pb}}}{dp_T d\eta}\Big/\frac{1}{d\sigma^{pp}/dp_T}\frac{d\sigma^{pp}}{dp_T d\eta}.\label{RAAdijet}
\end{equation}
nPDFs consist of 40 EPPS16 error sets of nuclear modification and 56 error sets of proton baseline, which are fully correlated with CT14 error sets.
The proton baseline largely cancels in the ratio, therefore we perform reweighting on EPPS16 error sets  only.
We combine the data points $y_a^\text{data}=R_{p\text{Pb}}^\text{norm}(p_T^\text{avg}, \eta_\text{dijet})$
from five  averaged dijet momentum bins $p_T^\text{avg}/\text{GeV}= [55,75], [75,95], [95,115], [115,150], [150,400]$ and from averaged dijet rapidity bins, which fall in the range $-3<\eta_\text{dijet}<3$. The reweighted uncertainties and new central value are found by \Eqs{eq:newerr} and \eq{eq:newmin}. In \Fig{fig5} we show the result for the lowest momentum bin (other $p_T^\text{avg}$ ranges not shown). 
We observe a large reduction in nPDF uncertainties as first reported in Ref.~\cite{Eskola:2019dui}. Importantly, the effect of this reweighting on other predictions can be obtained by replacing $D_{ak}$ and $y_a(\bm{z}_\text{min})$ in \Eqs{eq:newerr} and \eq{eq:newmin}.  The results for the nuclear modification factors in OO collisions  were shown in Figs.~2,3 and 4 in the main text.

LHC run 3 high statistics data of electroweak bosons and jet observables 
in  $p$Pb  collisions (where energy loss mechanisms are negligible) is expected to improve nPDF uncertainties~\cite{Citron:2018lsq}. OO and $p$O collisions could help to validate and improve nPDF fits at small nucleon number, but high collision energies.

\fakesection{$Z$-boson production in OO collisions}
The electroweak boson production in heavy-ion collisions have been used to access the initial state properties unobscured by the medium, e.g., to constrain the nPDFs. $Z$ bosons provide particularly clean experimental observables, which can be inferred from the di-lepton invariant mass spectrum. Therefore it is natural to expect that $Z$ bosons provide a high precision hard parton luminosity meter. Here we discuss the unexpected anti-correlation of nPDF uncertainties at different Bjorken-$x$ that makes this conclusion premature.

\begin{figure}
    \centering
\includegraphics[width=\linewidth]{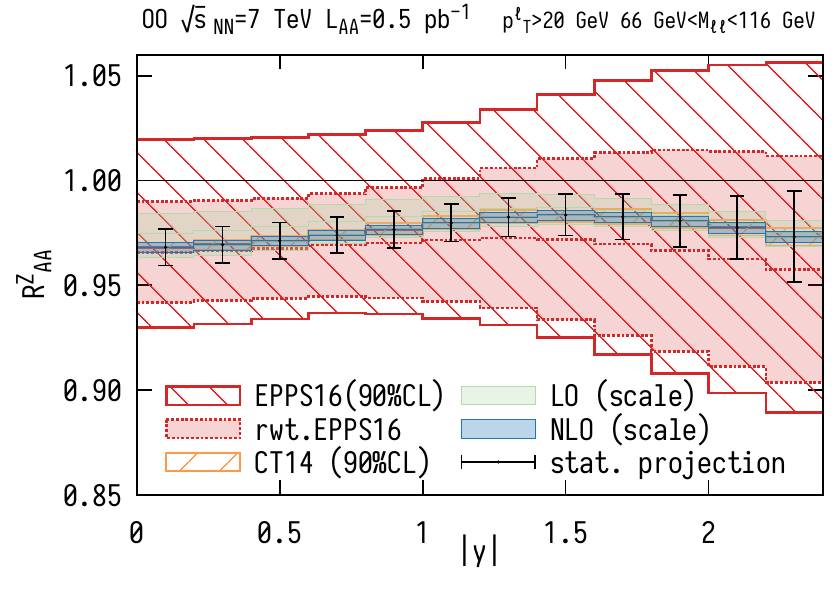}
    \caption{ The $Z$-boson nuclear modification factor.
     The red bands show nPDF uncertainties (reweighting is done
  by including additional $p$Pb dijet data). 
  The orange and blue bands show cancellation of fully correlated proton PDF and scale uncertainties.
    The error bars represent statistical uncertainties of mock data at 100\% efficiency at integrated luminosity of ${\mathcal L}_{AA}=0.5\,\text{pb}^{-1}$.
    }
    \label{fig6}
\end{figure}
We use the \texttt{NNLOJET} framework to calculate the $Z$-boson cross section at NLO in $pp$ and OO collisions at $\sqrt{s_{NN}}=7\,\text{TeV}$. In \Fig{fig6} we plot the $Z$ boson nuclear modification factor
\begin{align}
 R^{Z}_\text{AA, min bias}(|y|)= 
 \frac{1}{A^2} \frac{ {d\sigma^{Z}_\text{AA}}/{dy}}{ {d\sigma_{pp}^{Z}}/{dy}} 
\label{RAAZminbias}
\end{align}
as a function of absolute $Z$-boson rapidity. 
The theoretical uncertainties for differential $R_\text{AA}^{Z}$ range from 5\% to 9\%. We estimate that statistical uncertainties for the total sample of  $\mathcal{O}(10^5)$ $Z$ bosons in $-2.4<y<2.4$ range would be  $\mathcal{O}(1\%)$ for $R_\text{AA}^{Z}(|y|)$ shown in \Fig{fig6} and $\mathcal{O}(0.3\%)$ for the total fiducial cross section. This does not take into account other experimental uncertainties, in particular the luminosity normalization.

The total fiducial $Z$-boson cross section is used to normalize the jet nuclear modification factor and the result was shown in Fig.~4 in the main text. Contrary to initial expectations, the nPDF uncertainties do not cancel between $Z$-boson and jet cross sections. The origin of this can be traced back to the different Bjorken-$x$ regions of nPDFs that is probed by the two processes. 

Cross section predictions for hadron collisions $\sigma^{AB}$ can be computed through a convolution of the parton level cross section $\hat{\sigma}$ and
the parton luminosities given by the PDFs,
\begin{equation}
\label{eq:ABtojet}
\sigma^{AB} = \int dx_A d x_B f_{a}^{A}(x_A,\mu_F^2)f^{B}_{b}(x_B,\mu_F^2) \hat \sigma^{ab}.
\end{equation}
At leading order, the Bjorken-$x$ probed by $Z$ bosons and jets at the center of mass energy $s$ are given by
\begin{equation}
    x^{Z}_{A,B}=\frac{M_Z}{\sqrt{s}}e^{\pm y}, \quad x^{j}_{A,B}=\frac{p_T}{\sqrt{s}}( e^{\pm y_1}+ e^{\pm y_2}),
\end{equation}
where $M_Z$ and $y$ are the mass and rapidity of $Z$-boson, and $p_T$, $y_1$ and $y_2$ are the transverse momentum and rapidities of leading and subleading jets. We note that if $y=y_1=y_2$, then  $x^{Z}_{A,B}=x^{j}_{A,B}$ for $p_T = M_{Z}/2\approx 45\,\text{GeV}$, which corresponds to the lowest momentum bin in Fig.~4 in the main text.

\begin{figure}[t!]
    \centering
\includegraphics[width=\linewidth]{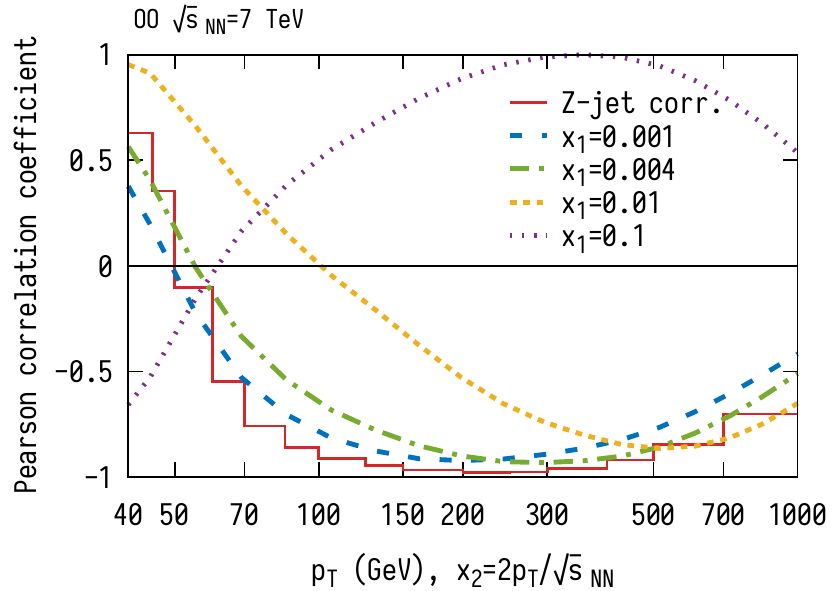}
    \caption{The correlation of nPDF uncertainties between inclusive $Z$-boson and jet cross sections as a function jet $p_T$. Also shown are nPDF uncertainty correlations of gluon nPDF $f_g(x_i,Q=M_Z)$ at $x_1$ and $x_2$, where $x_2=2p_T/\sqrt{s_{NN}}$.
    }
    \label{fig7}
\end{figure}

 In general, $x^{Z}_{A,B}\neq x^{j}_{A,B}$ and $Z$ bosons and jets are sensitive to partonic fluxes at different Bjorken-$x$. The uncertainties could still cancel if the nPDF error sets remains correlated over that $x$ range. We compute the correlation coefficient
\begin{equation}
    \text{Pearson corr. coef.} = \frac{\text{cov}(X,Y) }{\sqrt{\text{cov}(X,X)\text{cov}(Y,Y)}}
\end{equation}
between the total $Z$ boson~($X$) and inclusive jet cross section~($Y$)
evaluated on the 40 EPPS16 error sets. 
The result is shown as the red line in \Fig{fig7} for the NLO prediction, which is very similar to the correlation obtained at LO (not shown).
We observe positive cross section correlation for $p_T<50\,\text{GeV}$, which however turns negative at higher jet momentum.  For comparison, we plot the  correlation between gluon distribution functions $X=f_g(x_1,M_Z),Y=f_g(x_2,M_Z)$ for the same EPPS16 error sets.
We find that uncertainties of partons in the small $x$ shadowing region $x\ll 0.01$
are anti-correlated to those in the anti-shadowing region $x\approx 0.1$~\cite{Eskola:2016oht}.
The rapidity integrated cross sections are convolutions of products of PDFs at different Bjorken-$x$, but the observed correlation in cross sections follows closely that by partons at $x_1=0.004$ and $x_2=2p_T/\sqrt{s}$.

In summary, because of anti-correlation between parton fluxes probed by $Z$ bosons and jets, the theoretical nPDF uncertainties in the ratio of these fluxes add up instead of canceling. One way forward is simply to expect that with new data in global fits, the overall uncertainties will be sufficiently reduced. However, it would be also interesting to see if the present anti-correlation between nPDF uncertainties could be exploited  to increase the constraining power of such additional data.

\end{document}